\documentclass[twocolumn]{aastex61}
\pdfoutput=1 
\usepackage{ifthen, hyperref}
\usepackage{breakurl}
\usepackage{amsmath,amstext}
\usepackage[T1]{fontenc}
\usepackage{apjfonts} 
\usepackage[figure,figure*]{hypcap}
\usepackage{graphicx}
\usepackage[caption=false]{subfig}  
\usepackage{comment}
\usepackage{tabularx}
\usepackage{tablefootnote}

\shorttitle{Variability of RSGs in M31}
\shortauthors{Soraisam et al.}

\begin{document}

\title{Variability of Red Supergiants in M31 from the Palomar Transient Factory}

\author{Monika D.~Soraisam}
\affiliation{National Optical Astronomy Observatory, Tucson, AZ 85719, USA}
\author{Lars Bildsten}
\affiliation{Department of Physics, University of California, Santa Barbara, CA 93106, USA}
\affiliation{Kavli Institute for Theoretical Physics, University of California, Santa Barbara, CA 93106, USA}
\author{Maria R.~Drout}
\affiliation{Hubble and Carnegie-Dunlap Fellow}
\affiliation{The Observatories of the Carnegie Institution for Science, 813 Santa Barbara St., Pasadena, CA 91101, USA}
\author[0000-0002-4791-6724]{Evan B.~Bauer}
\affiliation{Department of Physics, University of California, Santa Barbara, CA 93106, USA}
\author{Marat Gilfanov}
\affiliation{Max Planck Institute for Astrophysics, Karl-Schwarzschild-Str.~1, 85748 Garching, Germany}
\affiliation{Space Research Institute, Russian Academy of Sciences, Profsoyuznaya~84/32, 117997 Moscow, Russia}
\author{Thomas Kupfer}
\affiliation{Division of Physics, Mathematics, and Astronomy, California Institute of Technology, Pasadena, CA 91125, USA}
\author{Russ R.~Laher}
\affiliation{Infrared Processing and Analysis Center, California Institute of Technology, Pasadena, CA 91125, USA}
\author{Frank Masci}
\affiliation{Infrared Processing and Analysis Center, California Institute of Technology, Pasadena, CA 91125, USA}
\author{Thomas A.~Prince}
\affiliation{Division of Physics, Mathematics, and Astronomy, California Institute of Technology, Pasadena, CA 91125, USA}
\author{Shrinivas~R.~Kulkarni}
\affiliation{Division of Physics, Mathematics, and Astronomy, California Institute of Technology, Pasadena, CA 91125, USA}
\author{Thomas Matheson}
\affiliation{National Optical Astronomy Observatory, Tucson, AZ 85719, USA}
\author{Abhijit Saha}
\affiliation{National Optical Astronomy Observatory, Tucson, AZ 85719, USA}
\correspondingauthor{Monika Soraisam}
\email{msoraisam@noao.edu}

\begin{abstract}

Most massive stars end their lives as Red Supergiants (RSGs), a short-lived evolution phase when they are known to pulsate with varying amplitudes. The RSG period-luminosity (PL) relation has been measured in the Milky Way, the Magellanic Clouds and M33 for about 120 stars in total. Using over 1500 epochs of $R$ band monitoring from the Palomar Transient Factory (PTF) survey over a five-year period, we study the variability of 255 spectroscopically cataloged RSGs in M31. We find that all RGSs brighter than $M_K\approx -10$~mag ($\log(L/L_\odot)>4.8$) are variable at $\Delta m_{R}>0.05$~mag. Our period analysis finds 63 with significant pulsation periods.  Using the periods found and the known values of $M_{K}$ for these stars, we derive the RSG PL relation in M31 and show that it is consistent with those derived earlier in other galaxies of different metallicities.  We also detect, for the first time, a sequence of likely first-overtone pulsations. Comparison to stellar evolution models from \texttt{MESA} confirms the first overtone hypothesis and indicates that the variable stars in this sample have $12~M_{\odot}<M<24~M_{\odot}$. As these RSGs are the immediate progenitors to Type II-P core-collapse supernovae (SNe), we also explore the implication of their variability in the initial-mass estimates for SN progenitors based on archival images of the progenitors. We find that this effect is small compared to the present measurement errors.

\end{abstract}

\keywords{groups: individual: M31 -- stars: massive  -- supergiants -- stars: oscillations -- surveys}

\section{Introduction}\label{intro}
Evolution of massive stars, those with zero age main sequence (ZAMS) spectral types O and B, populating the upper part of the Hertzsprung-Russell (HR) diagram, remains one of the most puzzling fields in stellar physics. Its details are influenced by many uncertain physical processes, which include, among others, mass loss, convection, and rotation (e.g., \citealt{Langer-1995}, \citealt{Maeder-2000}, \citealt{Mauron-2011}, \citealt{Georgy-2012}, \citealt{Beasor-2017}; see \citealt{Martins-2013} for a comparison of different models employing varied prescriptions for these physical processes). Furthermore, there are many unknowns surrounding the mapping of these stars to their final fates, with cases of observational results conflicting theory---e.g., what produces which types of core-collapse supernovae (SNe), and what results in black holes (\citealt{Heger-2003, Sukhbold-2016, Nathan-2011}; \citealt{Smartt-2}).

Despite the incomplete knowledge of these aspects of the evolution of massive stars, they remain from birth to death one of the key players moderating and tracing properties of galaxies, such as the star formation rates, energetics---both radiation and mechanical---and chemical enrichment (see \citealt{Massey-2013} for a recent review). Red Supergiants (RSGs), which are core He-burning (or beyond) stars representing the red-most excursion in the HR diagram during evolution of massive stars in the ZAMS mass range $\sim 10\mbox{--}30~M_{\odot}$ (for Population I stars), are the largest and among the optically brightest, and thus easily detected stars in the Local Universe. As a result of direct detections in pre-explosion images, they are recognized as the progenitors of the most abundant type of core-collapse SNe, namely the Type II-P \citep[e.g.,][]{Groh-2013, Van-Dyk-2003, Smartt-2009}. 

In addition, many RSGs are variable, the properties of which have long been interpreted as pulsations \citep[e.g.,][]{Stothers-1969, Guo-2002}. The driving mechanism for pulsation in RSGs is not yet fully understood but it is thought to be the $\kappa$ mechanism in the hydrogen ionization zone, coupled with convection, whose feedback however remains unknown \citep{Heger-1997,Yoon-2010}. Despite this theoretical uncertainty, the existence of an observed period-luminosity (PL) relation for RSGs entails their potential use as extragalactic distance indicators aided by their high luminosities and their higher prevalence as compared to Classical Cepheids \citep[e.g., ][]{Glass-1979, Feast-1980, Mould-1987}.

Most early work on the PL relation focused on the infrared, particularly the $K$ band, for determining the luminosity, as these objects suffer from strong absorption in the optical, for example due to presence of TiO in their atmospheres \citep{Pierce-2000}, and the bolometric correction is small in the $K$ band. From such a relation, \citet{Jurcevic-2000} were able to obtain a distance to the galaxy M101 consistent with the Cepheid distance.  However, studies investigating the PL relation for RSGs have been just a countable few \citep[e.g.,][]{Feast-1980,Wood-1983,Pierce-2000,Jurcevic-2000,Kiss-2006,Yang-2011,Yang-2012} in the almost four decades starting from the pioneering work of \citet{Glass-1979}.

RSGs have periods of typically several hundreds of days, and a semi-regular and complex variability; consequently, a sufficiently long baseline is needed for their studies. This has been largely made possible either through decades-long collections of the American Association of Variable Star Observers (e.g., \citealt{Kiss-2006} for the Galactic RSGs) or recent surveys such as ASAS \citep{Pojmanski-2002} and MACHO (\citealt{Alcock-1997}; e.g., \citealt{Yang-2011, Yang-2012} for the RSGs in the Magellanic Clouds). However, until now, there has not been a study of the variability of RSGs in the closest spiral galaxy to us, M31, even though a good number of them have been identified and cataloged (Sect.~\ref{m31_rsgs}). This was due to a dearth of {\em long-baseline time-domain} surveys for M31. However, with recent surveys such as the (intermediate) Palomar Transient Factory (iPTF) survey \citep{Law-2009,Rau-2009,Ofek-2012}, the WeCAPP survey \citep{Riffeser-2001} of the bulge of M31 (2000--2003; \citealt{Fliri-2006}) and the PAndromeda project (2010--2012) of Pan-STARRS 1 \citep{Lee-2012, Kaiser-2010}, such analyses are now possible. iPTF, in particular, possesses a baseline $\approx 1.5\mbox{--}2.5$ times longer than the other surveys and probes sources down to $m_{R}\approx 21$~mag over the whole galaxy (cf.~Sect.~\ref{iptf}). In this work, we present the study of the variability properties of RSGs in M31 and a measurement of their PL relation using observations from iPTF.

Furthermore, while the derivation of physical parameters for SN progenitors based on direct detections or upper limits in archival images has become more complex in recent years, such analyses still currently neglect any possible variability of the progenitor star \citep[e.g.,][]{Smartt-2, Davies-2017,Van-Dyk-2003}. Our knowledge of the behavior of the star in the few years leading to the supernova is far from complete, with observational and theoretical work pointing to the possibility of pre-explosion eruptions, \citep[e.g.,][]{Arnett-2014, Shiode-2014}, enhanced mass loss \citep[e.g.,][]{Morozova-2017,Beasor-2016,Beasor-2017}, and growth of pulsations \citep[e.g.,][]{Yoon-2010}. Even if pulsational variability similar to that observed in known RSG populations were still to be present during the archival imaging of the star, it could have a consequence in interpreting its initial mass. In particular, if pulsation amplitude is  luminosity-dependent, systematic effects could influence current observations, which point to a deficiency of higher-mass progenitors for SNe~II-P (dubbed the ``Red Supergiant Problem''; e.g., \citealt{Smartt-2009}; see Sect.~\ref{sne_mass}). As an illustrative example, we also assess quantitatively the possible extent of such an effect in this paper. 

The paper is organized as follows. In Sect.~\ref{m31_rsgs}, we introduce the sample of RSGs in M31. We describe the iPTF data in Sect.~\ref{iptf} and show the RSG lightcurves constructed from these data in Sect.~\ref{lcs} and the method to extract their periods in Sect.~\ref{period}. We then derive the PL relation in Sect.~\ref{PL} and compare it with theoretical \texttt{MESA} models in Sect.~\ref{mesa}, and touch upon the implication of the variability of RSGs for SN progenitors in Sect.~\ref{sne_mass}. We end with our conclusions in Sect.~\ref{conclude}.

\begin{figure*}[t]
\centering
\includegraphics[width=88mm]{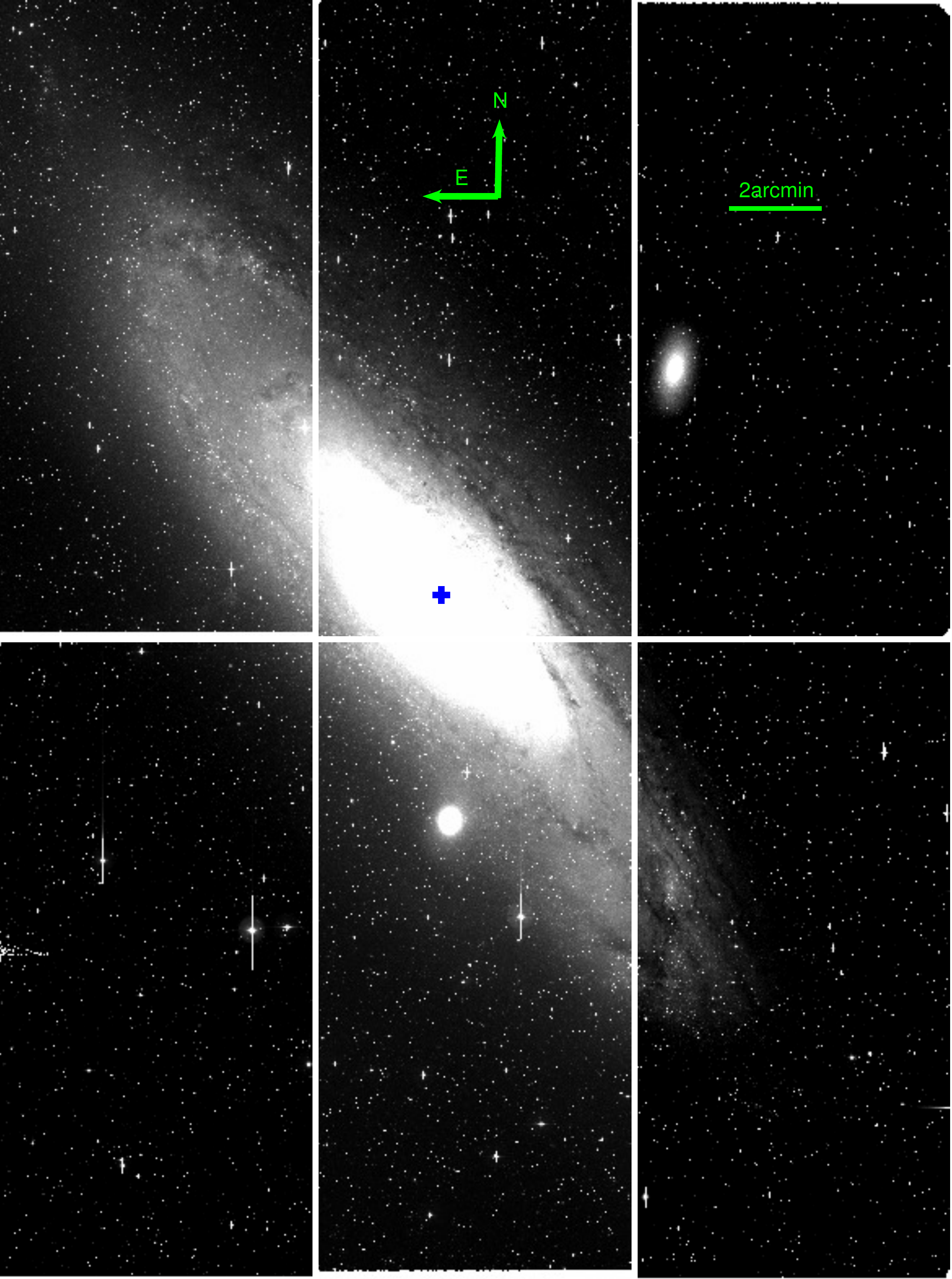}
\caption{iPTF footprint of M31 composed of six CCDs analyzed in this study. This composite image is made using the iPTF $R$ band reference image (a deep co-add of images) of each CCD. The blue cross marks the center of M31, while some of the masked regions (due to saturation, defective pixels, etc.) can be seen as bright lines scattered around in the various CCDs. \label{fig:iptf_footprint}}
\end{figure*}

\begin{figure*}[t]
\includegraphics[width=88mm]{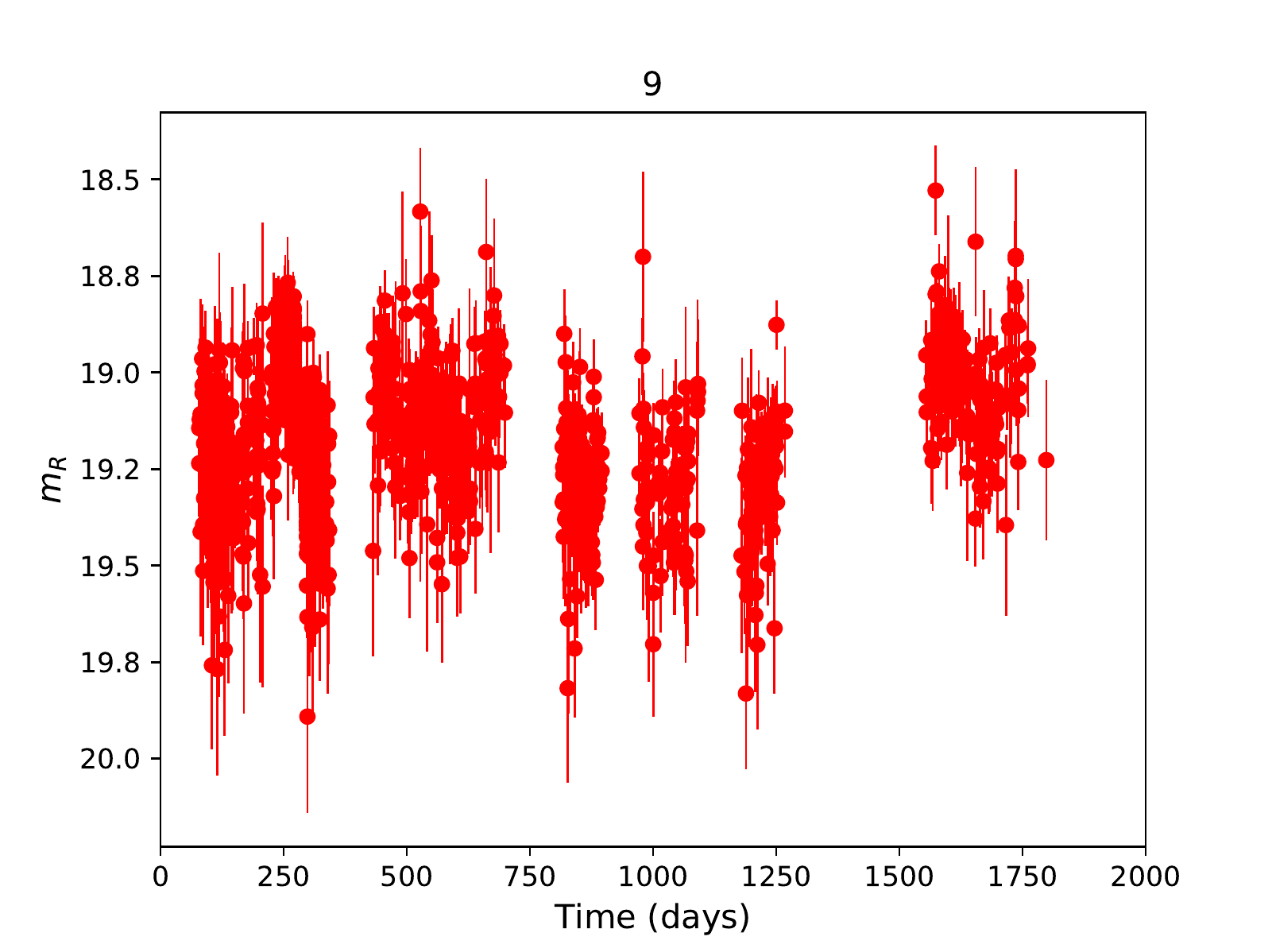}\hfil
\includegraphics[width=88mm]{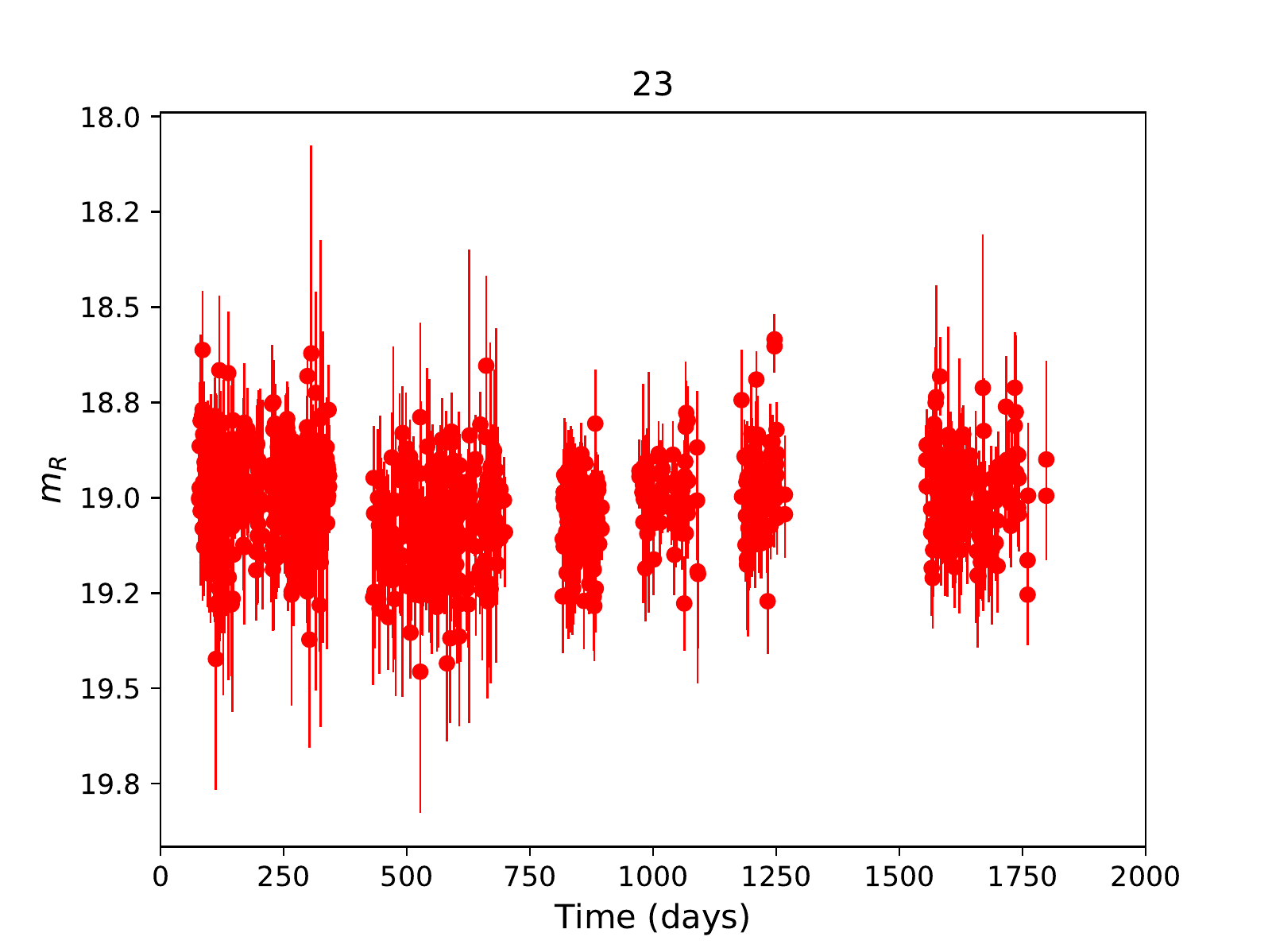}
\includegraphics[width=88mm]{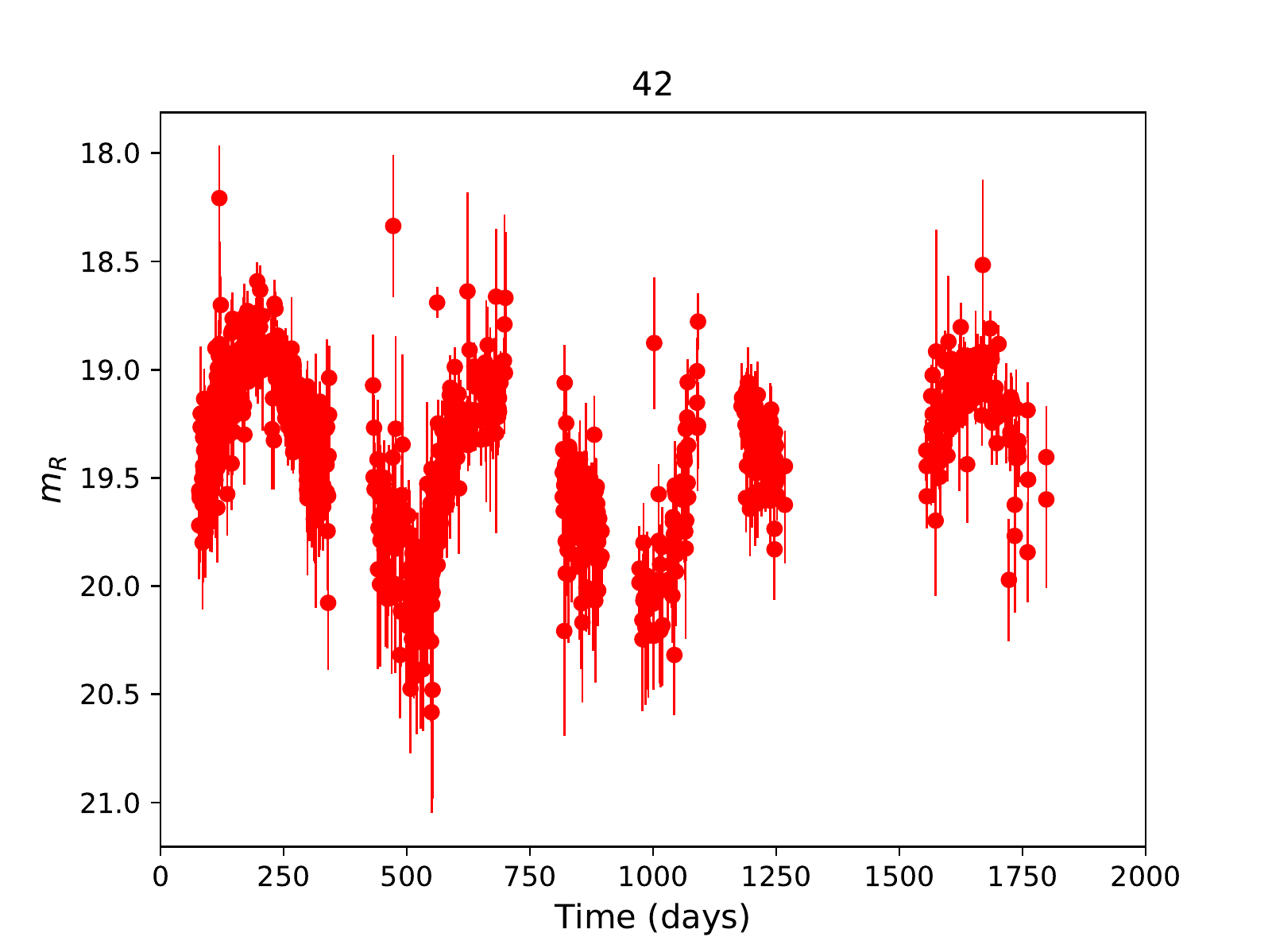}\hfil
\includegraphics[width=88mm]{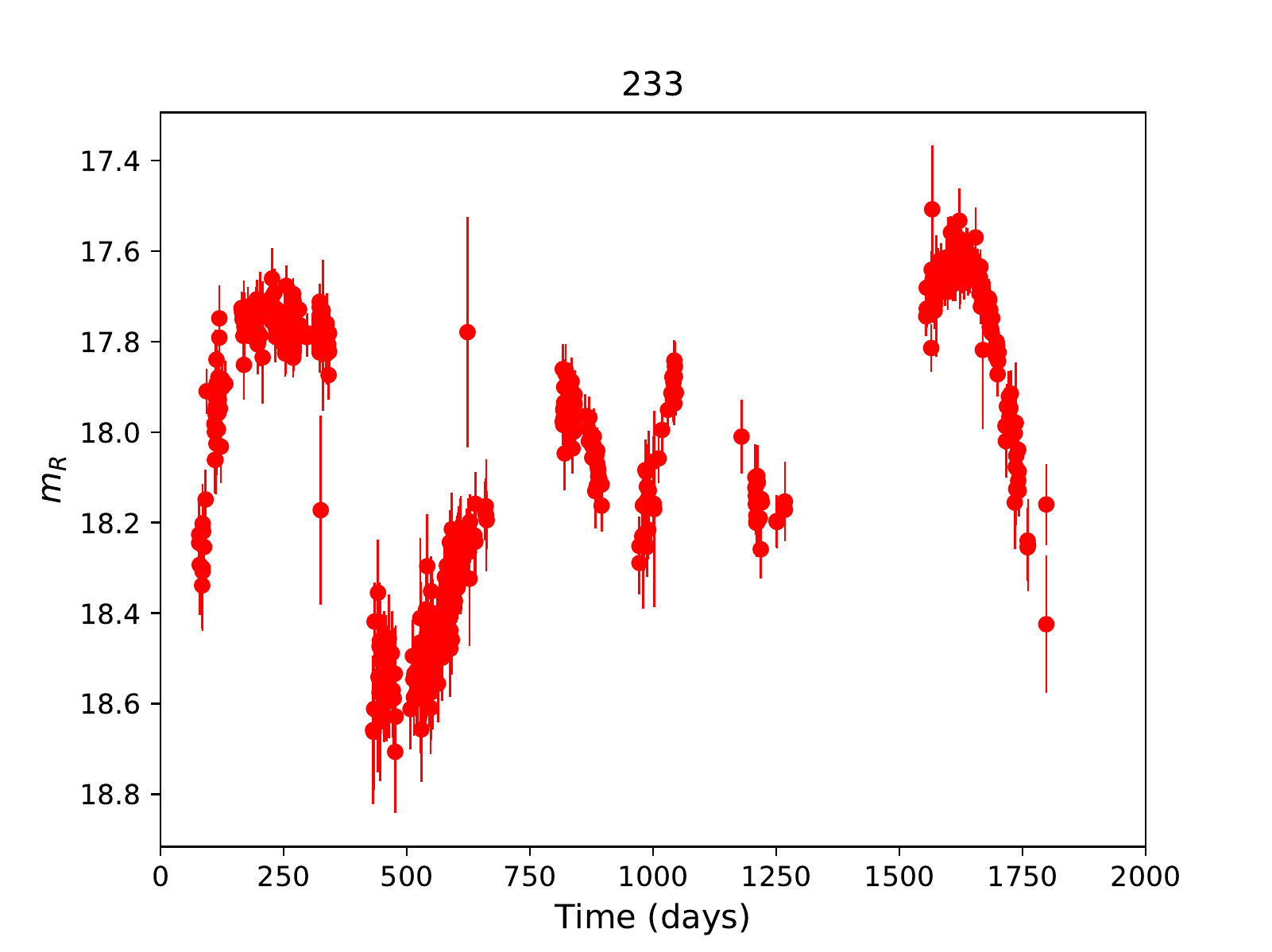}
\caption{Calibrated PTF lightcurves of RSGs in M31 (with IDs on top, that correspond to the order in which the RSGs appear in the ME16 catalog; cf.~Table~\ref{table_per}). The time axis is shown here with respect to a reference value of MJD 56000. These example lightcurves show the range of variabilities exhibited by the M31 RSGs in our sample. ID 23 is one RSG with its variability below our noise threshold (cf.~Fig.~\ref{fig:static}), while the rest are above this threshold.}\label{fig:lc_egs}
\end{figure*}

\section{Red Supergiants in M31}
\subsection{The sample of RSGs and their physical parameters}\label{m31_rsgs}

For studies of resolved stellar populations of several galaxies in the Local Group, the Local Group Galaxies Survey (LGGS; \citealt{Massey-2006, Massey-2007}) has proved to be a milestone. The survey was conducted with the 4-meter telescopes at the Kitt Peak National Observatory (KPNO) and the Cerro Tololo Inter-American Observatory. In particular, 10 fields of M31 covering $2.2~\rm{deg}^{2}$, were imaged with the Mosaic CCD camera at the 4-meter Mayall telescope at KPNO, from which broadband photometric measurements in multiple optical filters ($U,B,V,R,I$) of over 350,000 stars in M31 have been published by \citet{Massey-2006}. They achieved a photometric precision of $\approx (1\mbox{--}2)\%$ at 21~mag, thus providing an excellent catalog, particularly for studies of massive stars.   

\citet{Massey-1998} formulated a method for photometrically selecting RSGs, which serves to remove as far as possible the contaminating foreground red dwarfs. The author used the two-color diagram of $B-V$ versus $V-R$, wherein the two classes of objects are found to separate out into two sequences. For these spectral types, the $V-R$ color mostly traces the effective temperature ($T_{\rm eff}$) while the $B-V$ color also traces the surface gravity. Lower surface gravity objects (i.e., RSGs) have significantly redder $B-V$ values (by some tenths of a magnitude) due to increased importance of metal line blanketing, which is prominently expressed in the $B$ filter. Using the LGGS catalog, \citet{Massey-2009} compiled a sample of 437 photometrically selected RSG candidates based on these two colors. From this photometric sample, \citet{Massey-2016} [hereafter ME16] then obtained spectra of 255 stars and measured their radial velocities, in order to confirm their membership of M31. This yielded a large, spectroscopically pure sample of RSGs, which forms the fiducial sample of RSGs for our study, coupled with the data from the iPTF survey (described below). 

For their sample of confirmed RSGs, ME16 derived physical parameters (e.g., luminosities, temperatures), which we utilize in the sections below (cf.~Sect.~\ref{PL}). In order to derive luminosities for these stars, ME16 used available near-IR $K$ band photometry, in order to take advantage of the small bolometric and extinction corrections as compared to the optical bands. $K$ band photometric measurements were acquired from a combination of targeted observations (with FLAMINGOS; \citealt{Massey-2009}) and the Two Micron All-Sky Survey (2MASS; \citealt{Cutri-2003}). The bolometric correction for $K$ band was computed as a function of $T_{\rm eff}$, and values of $T_{\rm eff}$ for the stars were determined by fitting the \texttt{MARCS} atmospheric models to the observed spectra.  Finally, ME16 also obtained initial-mass estimates of the stars by comparing to the \texttt{GENEVA} stellar evolutionary tracks (see ME16 for more details), and spectral types following \citet{Levesque-2005} using the strengths of TiO bands (late K and M type stars) and that of G band and Ca~I $\lambda 4226$ (early and mid-K type stars).

\citet{Davies-2013} contend that the $T_{\rm eff}$ values obtained from fitting the TiO bands in optical spectra---the method employed by ME16---underestimate the true values by several hundred Kelvin, since these molecular lines form high up in the atmosphere with lower temperature. In contrast, \citet{Davies-2013} find $T_{\rm eff}\approx4150\pm150$~K for all spectral types of RSGs in the Magellanic Clouds that they have analyzed, when fitting the continuum of the full spectral energy distribution. However, \citet{Massey-2017} argue that this uniformity in $T_{\rm eff}$ is inconsistent with the observed variation of RSG spectral types with metallicity (for example, comparing the Milky way, LMC and SMC RSGs; \citealt{Levesque-2006}), and the fact that the Hayashi limit moves to higher $T_{\rm eff}$ values with decrease in metallicity \citep[e.g.,][]{Sugimoto, Chun17}. 
Thus, the uncertainty in the temperature scale of RSGs is as yet unresolved.   
Indeed, the extent of the atmospheres of these supergiants and its consequences for the occurrence of convection, stratification in the temperature, etc., undoubtedly add to the complexities in modeling them, which still remains a challenge. For this paper, we mostly use the $K$-band magnitudes of these sources from ME16, and turn to their luminosities for comparing to stellar evolutionary models (Sect.~\ref{mesa}) and in discussing the SN progenitor mass estimates (Sect.~\ref{sne_mass}).

\subsection{The iPTF data}\label{iptf}
The iPTF survey, carried out with the 1.2-meter Samuel Oschin telescope at Palomar Observatory, continued imaging M31 (including the outskirts of this galaxy) from its predecessor, the PTF survey \citep{Law-2009, Rau-2009, Ofek-2012}, with the same wide-field detector covering more than $7~{\rm deg}^2$ with 7 active CCDs, and a typical spatial sampling of $2''$ full-width at half maximum (FWHM). The baseline of the data for M31 analyzed here extends from 2012 May until 2017 February. The imaging was done in two passbands, $R$ and $g$, with the largest fraction ($>80\%$) in $R$ at a limiting magnitude reaching $\approx 21$~mag and an average cadence of 1 day (typically taking 2 observations per night).  

We thus use only the $R$ band PTF/iPTF (hereafter simply PTF) data set of about 10000 images. This set comprises over 1500 observation epochs of M31 covering $1.8\times2.4~{\rm deg}^{2}$ with 6 CCDs (as shown in Fig.~\ref{fig:iptf_footprint}), over the approximately 5 years of baseline. Difference imaging was performed on all these images via the difference imaging pipeline of the PTF collaboration, the IPAC/iPTF Discovery Engine (PTFIDE; \citealt{Masci-2016}), and this pipeline also carried out detections of varying sources on the difference images, which we term here as ``raw detections''.

\subsection{Optical lightcurves of RSGs and their variability}\label{lcs}

The PTFIDE raw detections catalog, in principle, provides us with a means to determine whether a source in M31 is variable or not. In fact, we find from the results of cross-matching (with a search radius of $2''$, consistent with the typical FWHM) that all RSGs lying within the PTF footprint (253) are flagged as variable by the PTFIDE pipeline. However, as already established by \citet{Masci-2016} and \cite{Soraisam-2017}, these ``raw detections'' are largely dominated by artifacts of image differencing, for example, dipoles from imperfect PSF-matching, edges of masked image regions due to saturated stars, CCD defects, etc. 

For our science goal, we do not want to compromise completeness by resorting to thresholding on output parameters of image differencing. Rather, we implement forced photometry to construct the RSG lightcurves, using the difference images themselves, and analyze their variability. Difference imaging significantly alleviates the problem of crowding in the M31 fields for measuring their fluxes. Subtraction artifacts in individual difference images will largely be reflected as rogue points in the lightcurves, which we effectively deal with (e.g., by masking those points) without discarding the source. 

In the same manner as \citet{Soraisam-2017}, we perform aperture photometry at the positions of the M31 RSGs and apply a curve-of-growth correction to the measured fluxes. The latter is a different factor for each star applied to correct for the flux missed due to the limited size of the aperture (taken to be the FWHM). The subtracted fluxes of the sources are then added using the template/reference image PSF-fit photometry catalog and calibrated using the relevant zero-point in the science images (see \citealt{Masci-2016} for details). We drop 9 RSGs with bad photometry in a large section of their lightcurves, for example due to their location close to masked parts of images or in regions close to the bulge where the quality of image differencing is poor. Example calibrated lightcurves are shown in Fig.~\ref{fig:lc_egs}\footnote{The PTF lightcurves of all the RSGs studied here are available from the corresponding author.}.

\begin{figure}[t]
\centering
\includegraphics[width=88mm]{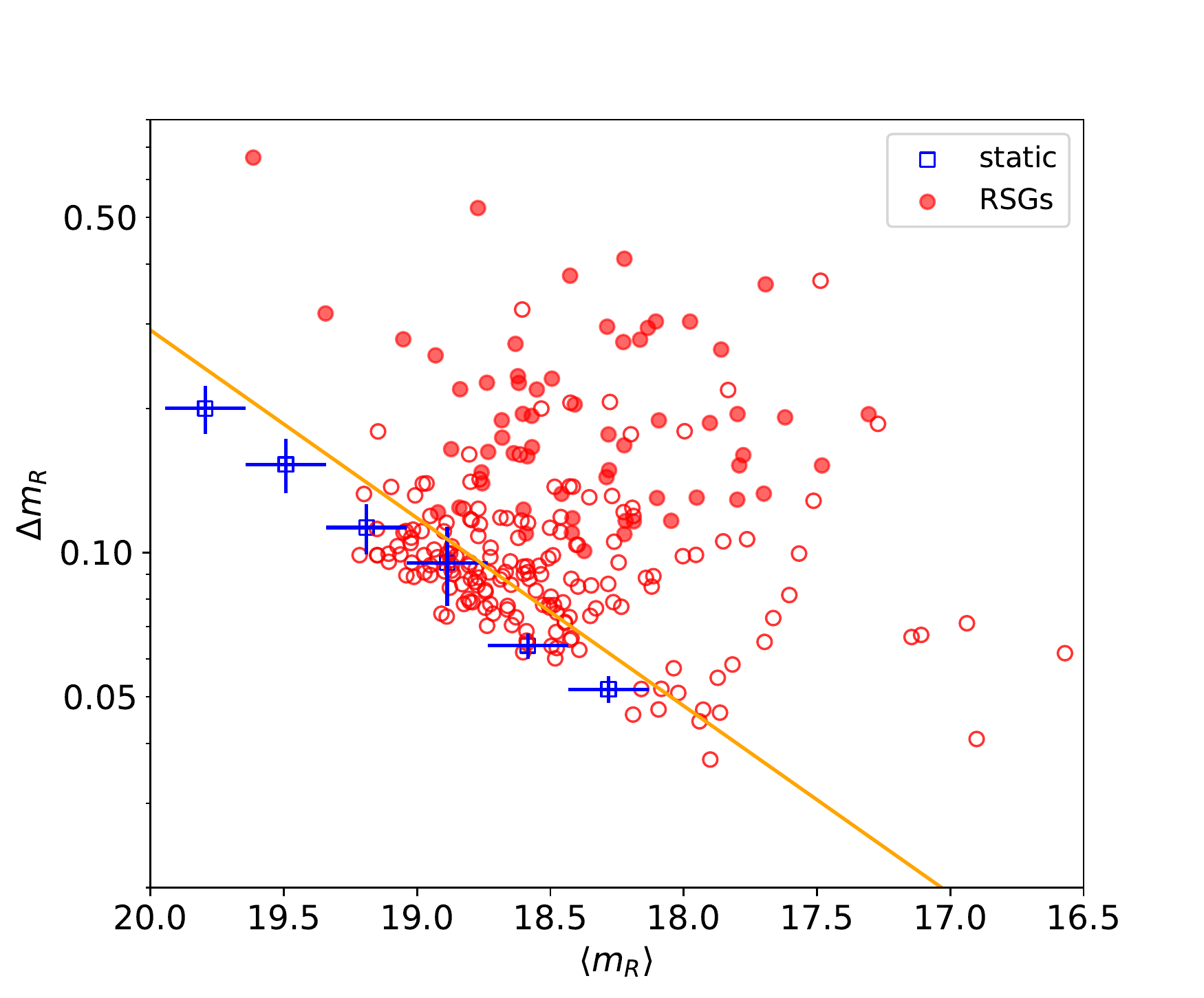}
\caption{RMS deviation vs average magnitude of lightcurve for static stars (see text for definition) shown in blue and {\em individual} RSGs shown as red circles. The orange line is drawn visually to demarcate the regions where the RSG variability sufficiently exceeds the noise as quantified via the RMS measurements for static stars. The periodicity analysis is performed only for RSGs above the orange line and the filled circles indicate those for which significant periods are found (Sect.~\ref{period}).}\label{fig:static}
\end{figure}

To examine the variability of these RSGs over the full baseline, we compute the root-mean-square (RMS) deviations from the mean for all of the extracted lightcurves, and compare them to those of {\em static} stars. We define a static star as a star in the catalog of the reference image used in the image subtraction, having no detection from the PTFIDE pipeline in all of the input images used in the subtraction (also termed science images by \citealt{Masci-2016}). We extract the lightcurves for these static stars in the same manner as those of the RSGs. Since the surface brightness of M31 is not uniform over the face of the whole galaxy, there is a gradient in the local background noise of a resulting difference image, which deteriorates severely particularly toward the bulge.  The RSGs, on the other hand, are largely distributed in the disk of the galaxy. To obtain a sample of static stars with a similar background noise distribution as the population of RSGs, we select the static stars that are located in proximity (within about $1'$) of any of the RSGs.

In Fig.~\ref{fig:static} we plot the distribution of the RMS deviations found for these static stars as a function of their average $R$ band magnitudes over the baseline of the survey (blue points). RMS values increase toward fainter magnitudes, and the distribution we find for stars in M31 agrees well with that for M-dwarf stars in PTF derived by \citet{Law-2012}. Also shown in Fig.~\ref{fig:static} are the M31 RSGs (red circles) and we see that the variabilities (as determined by the RMS deviations) of some of the RSGs, around the mean $\left<m_{R}\right>$ range of 18.0--19.0, are comparable to those of static stars, i.e.,  their variability is consistent with noise. Filled circles indicate RSGs with measured periods (Sect.~\ref{period}).  

We visually separate the RSGs with significant variability from the RSGs whose RMS deviations are consistent with noise. To this end, we draw a straight line in the plane of $\left<m_{R}\right>\mbox{--}\Delta m_{R}$, with the slope similar to the linear fit through the static stars and try different intercepts to most efficiently select significantly variable RGSs. This demarcation is shown with the orange line in Fig.~\ref{fig:static}. This method is not very stringent, but we verified that the main results of this paper remain the same with small shifts in this line. Adopting this demarcation, we find that 167 RSGs, constituting $\approx 70\%$ of our sample, have measurable variability from PTF. These are examined in the following section to determine what periodicity, if any, is present in their multi-year light curves.

\begin{figure*}[t]
\centering
\includegraphics[width=88mm]{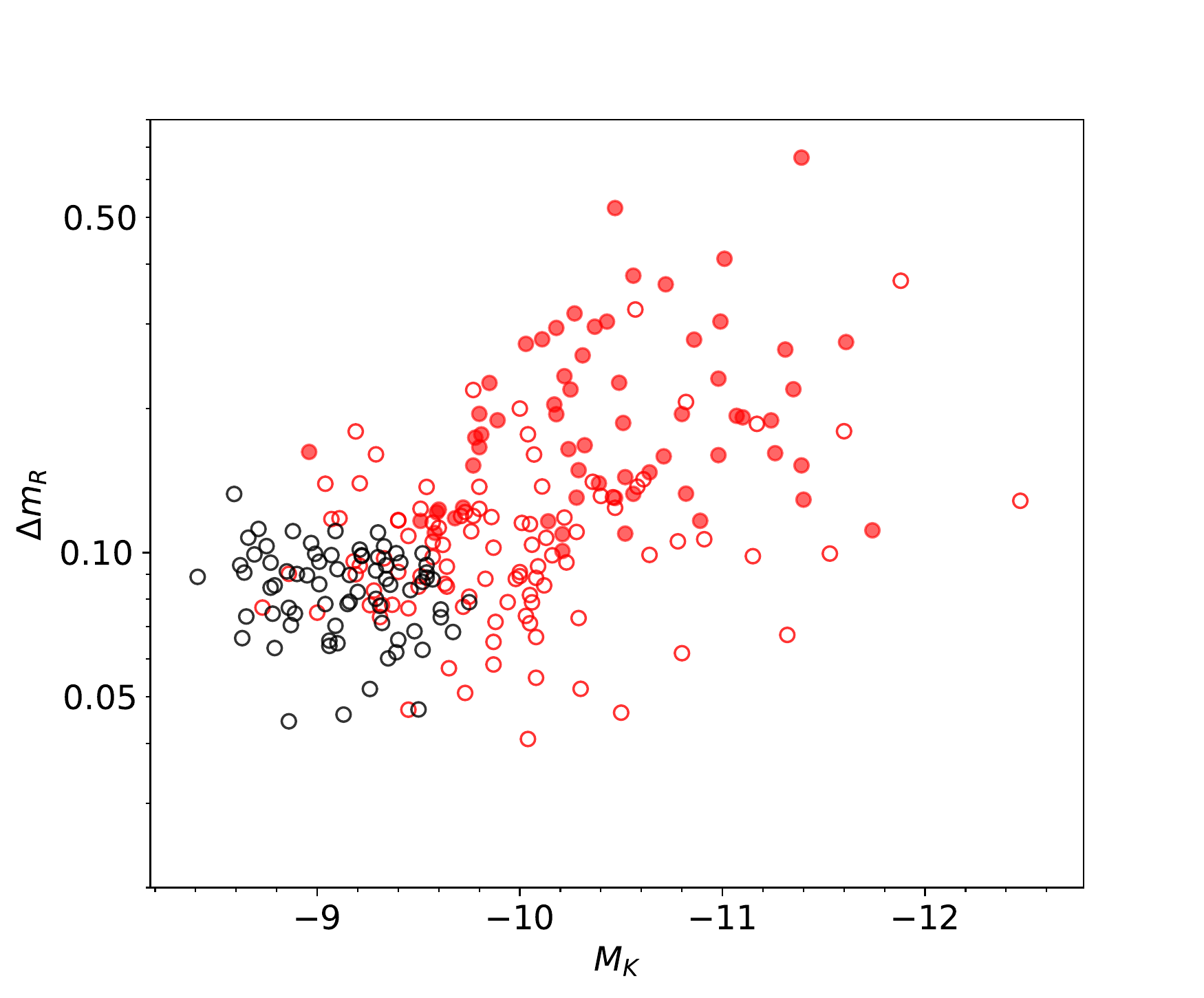}\hfil\includegraphics[width=88mm]{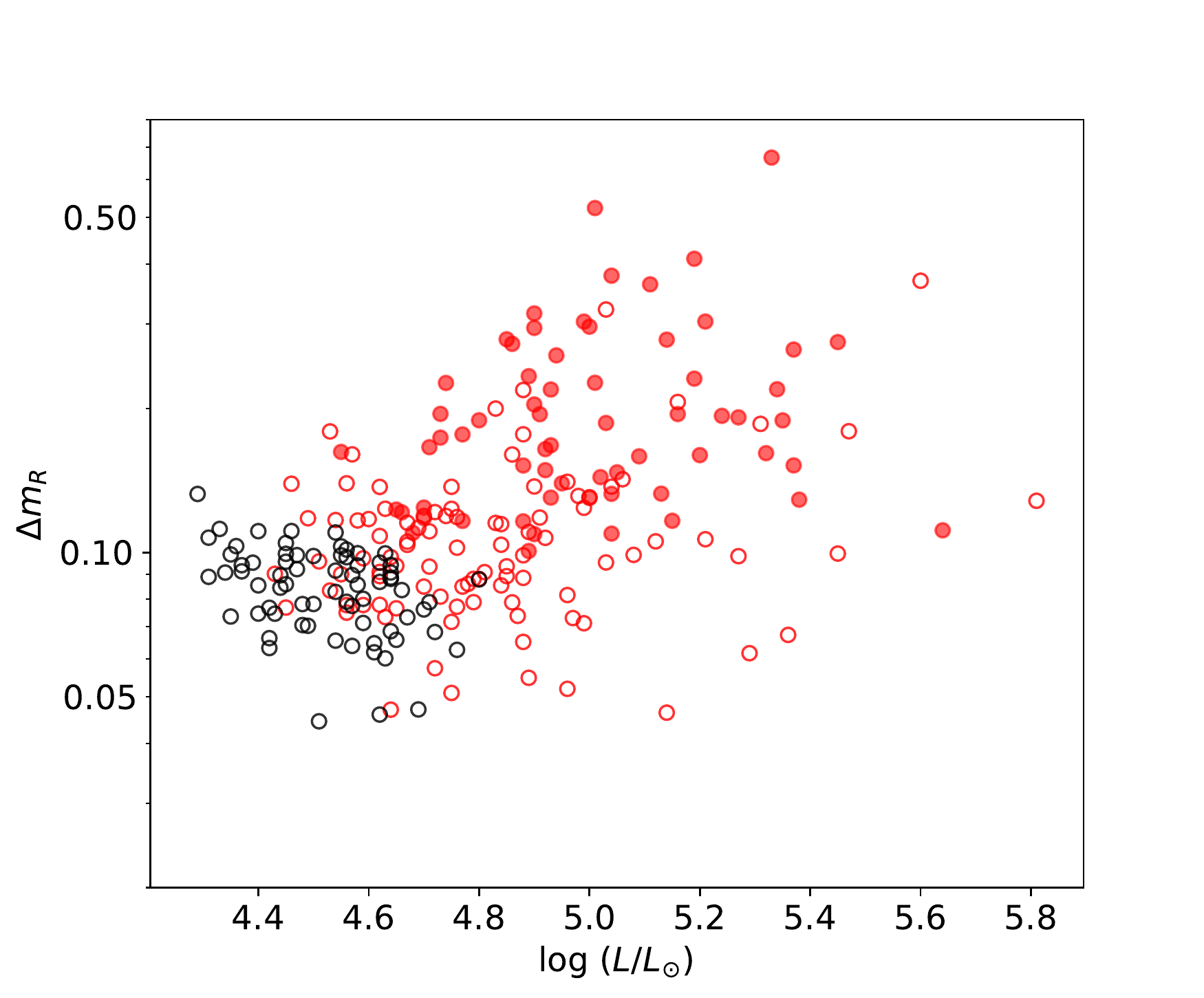}
\caption{RMS deviations for the RSG lightcurves against absolute $K$ magnitudes ({\it left}) and luminosities ({\it right}), both obtained from ME16. The red circles represent those with variabilities greater than that of static stars, with the filled ones indicating those for which we find significant periods. The black circles represent RSGs having RMS values consistent with noise (cf.~Fig.~\ref{fig:static}).}
\label{fig:Lrms}
\end{figure*}

In Fig.~\ref{fig:Lrms}, we plot the RMS deviations for our sample of M31 RSGs versus their $M_{K}$ magnitudes (left panel) and luminosities (right panel), whenever available from ME16. RSGs with significant variability are plotted in red, while those with PTF upper limits on variability (those below the orange line in Fig.~\ref{fig:static}) are plotted in black. 

As can be seen, at low luminosities there is an overlap between the RSGs with detected variability and those below our detection threshold, despite the fact that they fully separate in the $\left<m_{R}\right>\mbox{--}\Delta m_{R}$ plane. This can be attributed to the varying degree of extinction in the $R$ band suffered by individual RSGs, which subsequently modulates their signal-to-noise ratio (S/N). However, despite this overlap, it is notable that all of the ``non-varying'' RSGs are characterized by low intrinsic luminosity. All RSGs in our sample brighter than $M_{K}\approx-10$ (and $\log L/L_{\odot}\approx4.8$) show variability and many of them at the level $\Delta m_{R}>0.1$. We assume this variability of RSGs to be associated with pulsations \citep[e.g.][]{Heger-1997, Guo-2002}.
Further, our result points to variability/pulsations occurring already for  $\log L/L_{\odot}\gtrsim4.6$. Theoretical calculations of \citet{Yoon-2010} could only find pulsations above $\log L/L_{\odot}>4.95$ (e.g., their Fig.~1). This is, however, due to the fact that the implicit scheme adopted in stellar evolution calculations introduces substantial damping and thus, they could only place a lower limit on the existence of pulsations in RSGs{\footnote{Private communication with M.~Cantiello.}}. Our result on the other hand, shows that stars with mass lower than the $16\mbox{--}17~M_{\odot}$ limit found by \citet{Yoon-2010}, can also pulsate.
In addition, we find the hint of a positive correlation between variability amplitude and brightness for the RSGs with detected variability. In Sect.~\ref{sne_mass}, we will explore the influence that this observed variability can have on progenitor mass estimates of core collapse supernovae, in particular the trend of larger amplitude variations for more luminous RSGs.

\begin{figure*}[t]
\includegraphics[width=\textwidth, height=60mm]{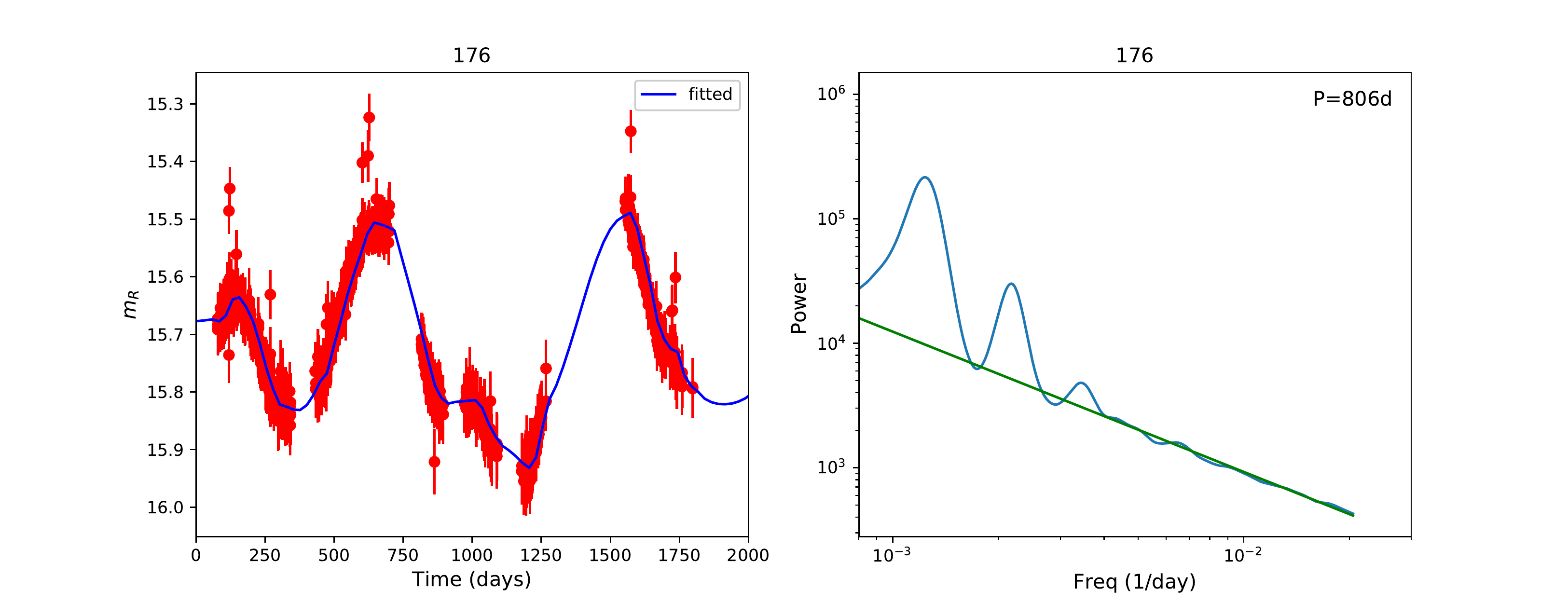}
\includegraphics[width=\textwidth, height=60mm]{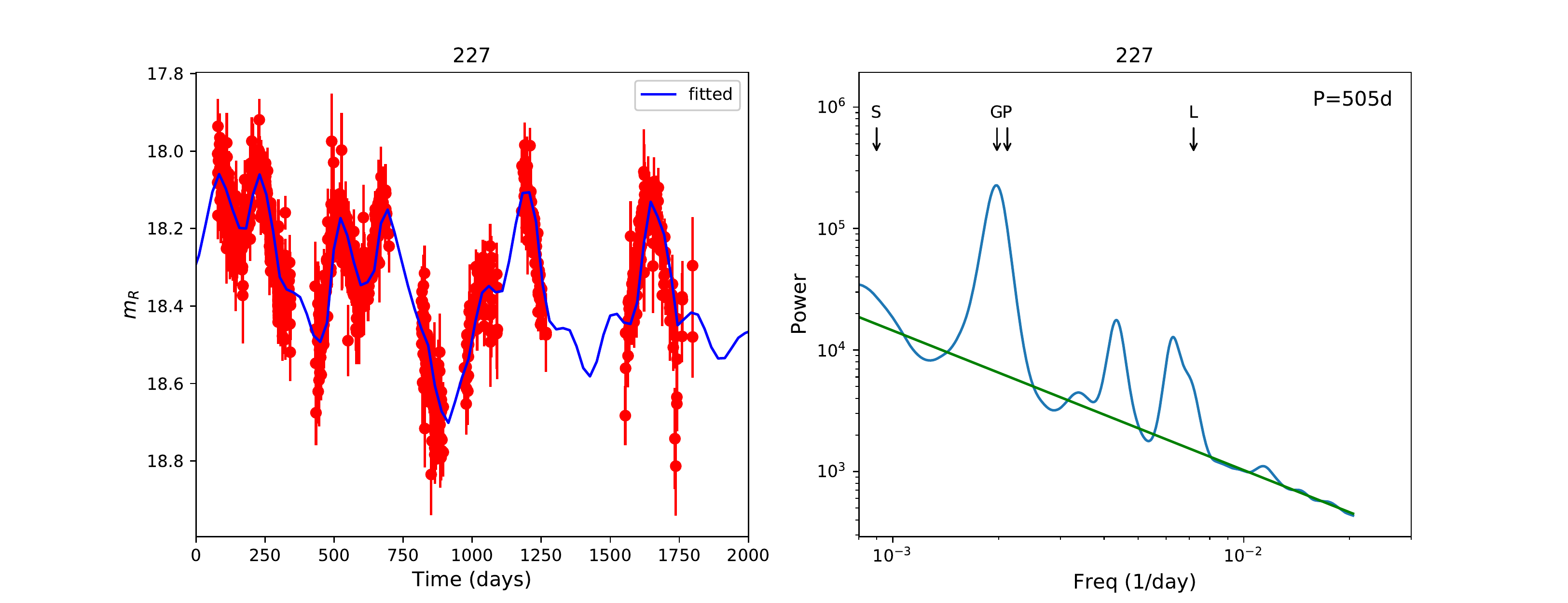}
\includegraphics[width=\textwidth, height=60mm]{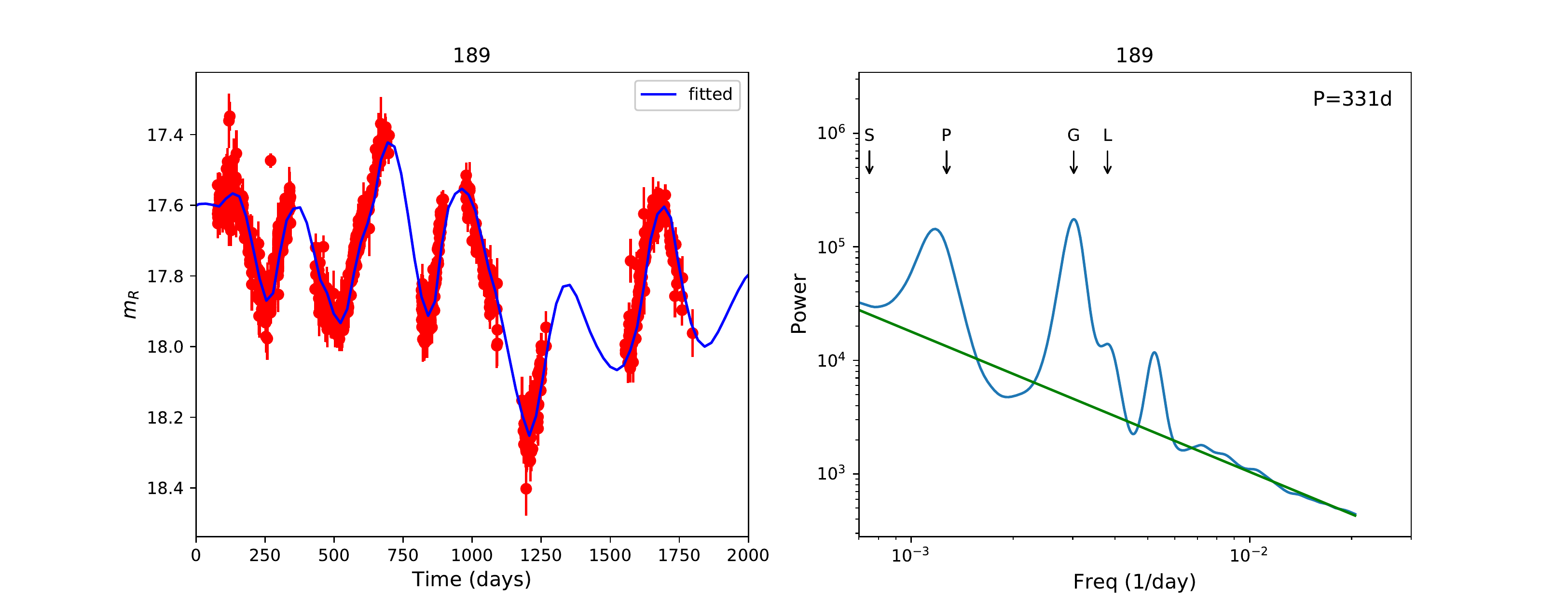}
\caption{{\it Left}: calibrated PTF lightcurves of RSGs in M31 (with their IDs on top, cf.~Table~\ref{table_per}); the time axis is with respect to a reference value of MJD 56000 and the blue curve is the fitted model from the Gaussian Process regression (see Sect.~\ref{period}). {\it Right}: corresponding power spectra of the lightcurves, constructed via the method detailed in Sect.~\ref{period}. The green line is a fit to the background red-noise. Period corresponding to the dominant mode in the power spectrum, if any,  is indicated in the legend. The arrows in the panels for ID 227 and 189 indicate dominant frequencies found by different methods---Lomb Scargle (L), supersmoother (S), phase dispersion minimization (P), this work (G)---(see Sect.~\ref{period}).}\label{fig:lc_spec_1}
\end{figure*}

\begin{figure*}
\ContinuedFloat
\includegraphics[width=\textwidth, height=60mm]{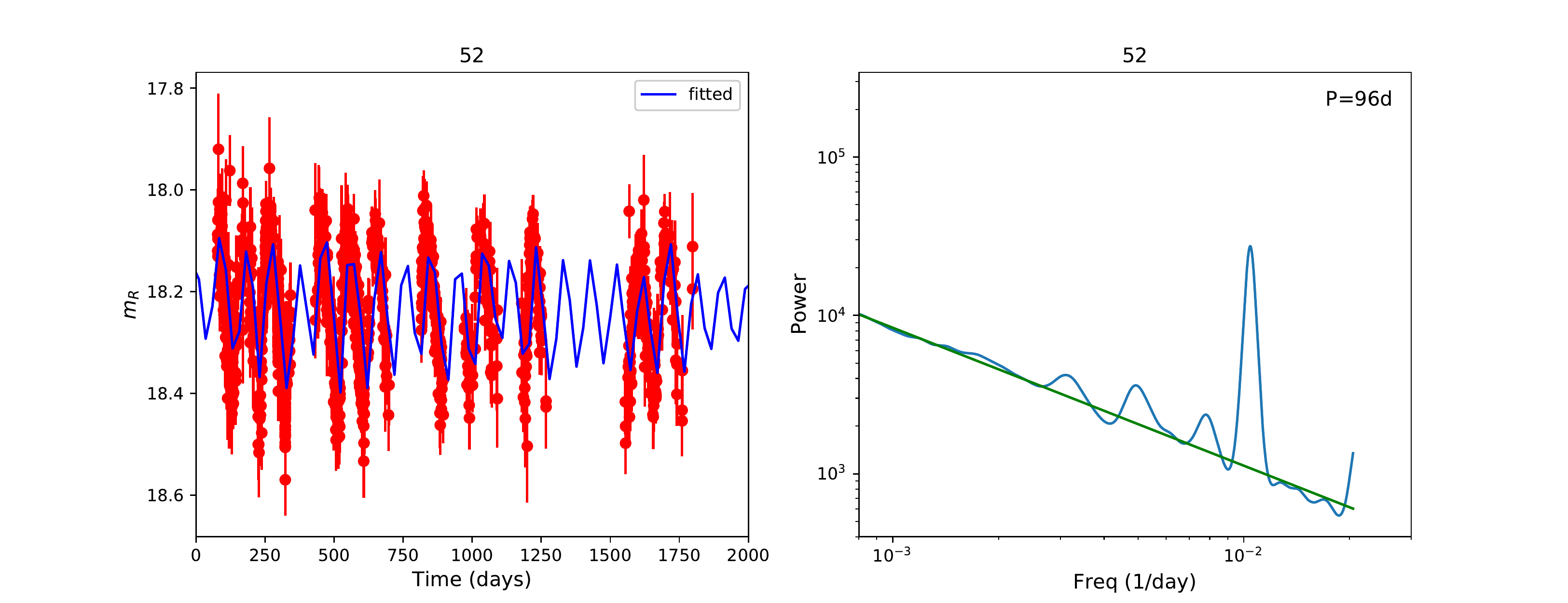}
\includegraphics[width=\textwidth, height=60mm]{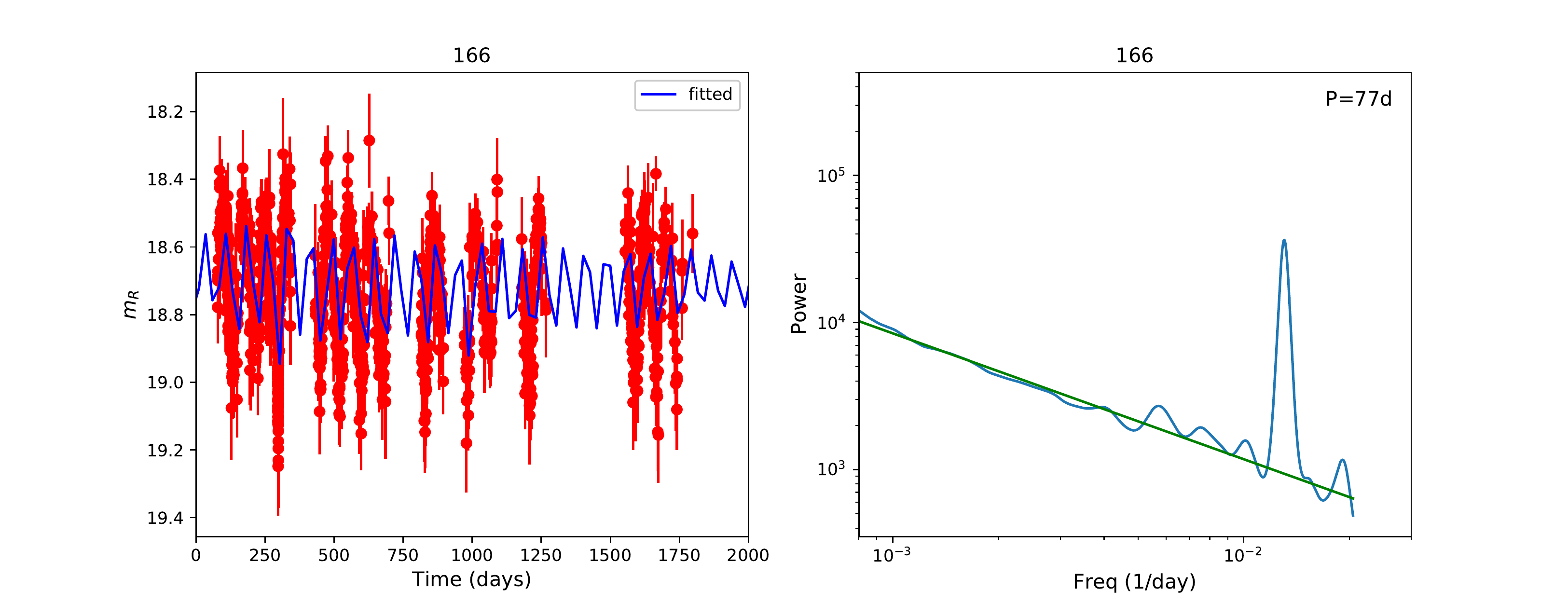}
\includegraphics[width=\textwidth, height=60mm]{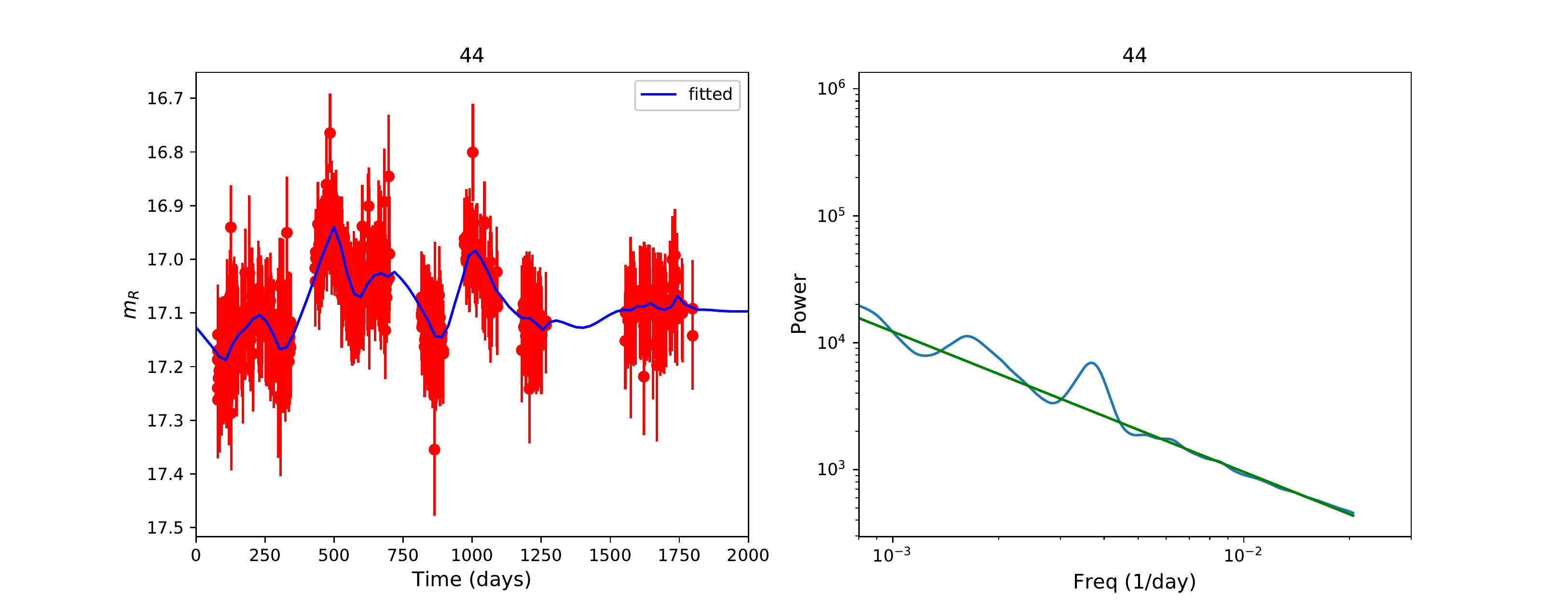}
\includegraphics[width=\textwidth, height=60mm]{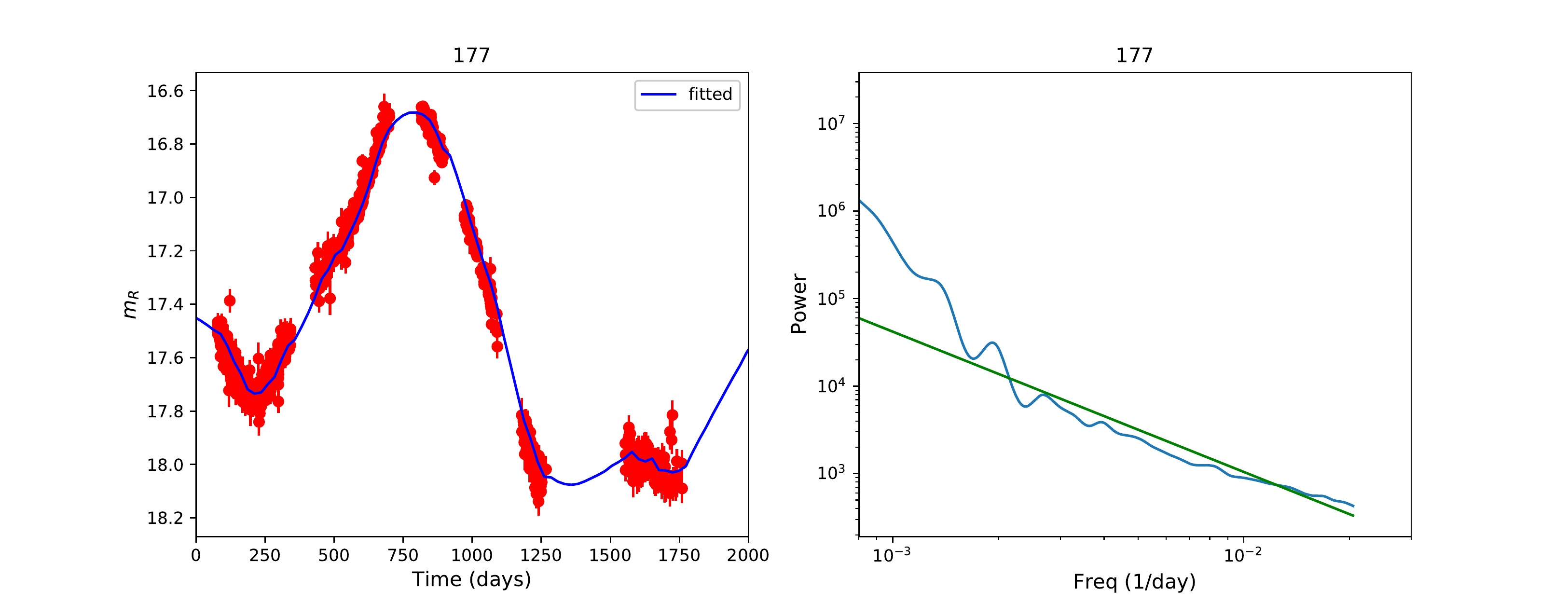}
\caption{continued. The last two panels show examples of RSGs, where we do not find any significant period.}
\end{figure*}

\subsection{Period determination}\label{period}

The lightcurves of RSGs are known to be semi-regular, and some of them are completely irregular (see Fig.~\ref{fig:lc_egs}). The periodicity of unevenly-sampled {\em regular} time-series/lightcurves has been efficiently handled by conventional standard algorithms, for example, Lomb-Scargle periodogram ( LS; \citealt{Lomb-1976, Scargle-1982}), phase dispersion minimization (PDM; \citealt{Stellingwerf-1978}), analysis of variance \citep{Schwarzenberg-1989}, and the hybrid algorithm of \citet{Saha-2017}.  However, for the RSGs, these methods tend to fall short, as these objects are characterized by one of the most complex lightcurve morphologies. This includes a strong red-noise component speculated to arise from the convective cells on the surface and multi-periodicity. For example, \citet{Kiss-2006} found two distinct periods for a significant fraction of their Galactic RSG sample over a baseline extending to more than 5 decades: one on the order of a few hundred days, which is typically associated with RSG pulsations, and the other greater than 1000 days, similar to the long secondary periods in Miras with nature unknown (see also \citealt{Percy-2013}). Compared to this, our baseline will not be sensitive to pick up these longer ($>1000$~days) periods.

\begin{figure}[t]
\centering
\includegraphics[width=88mm]{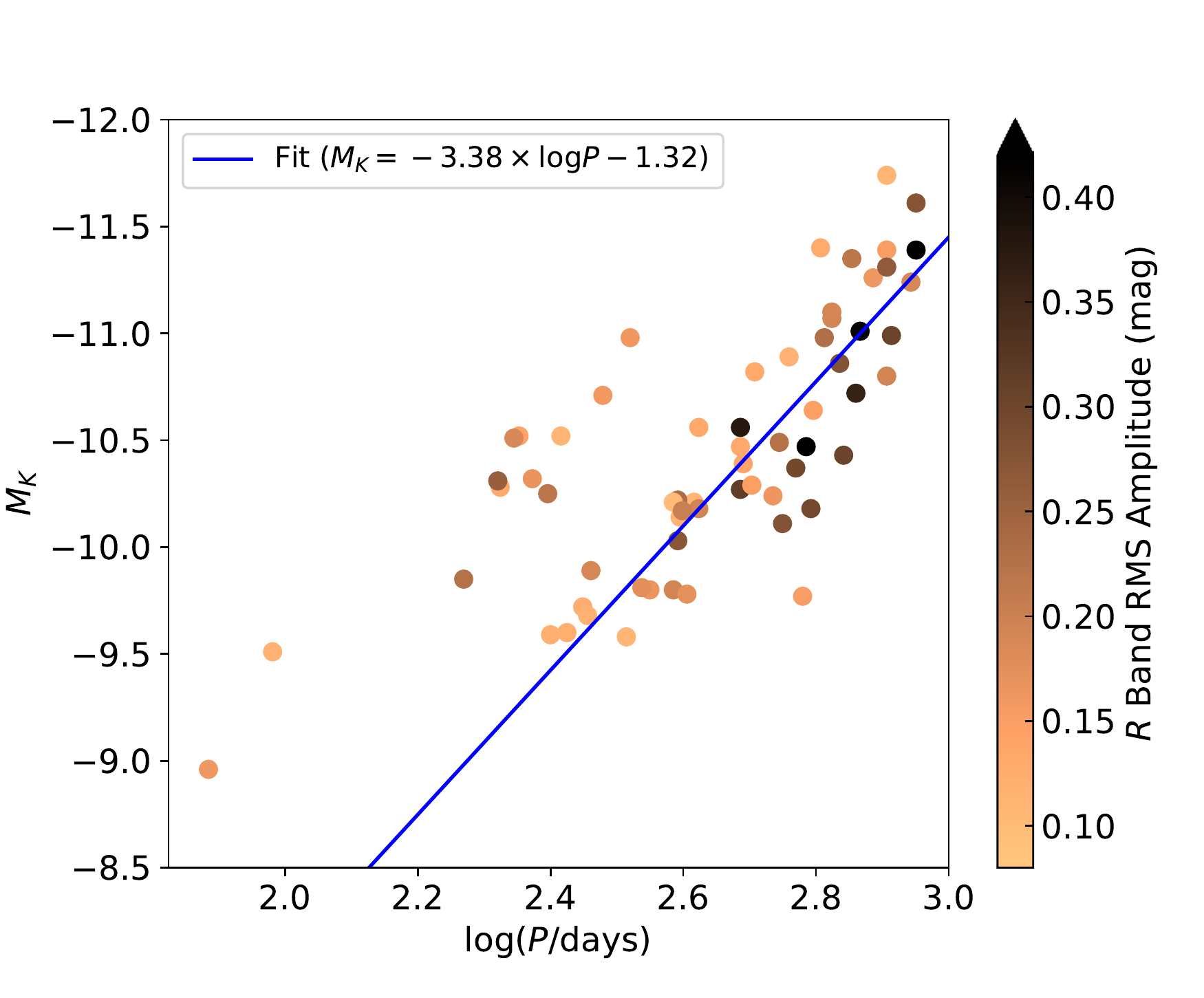}
\caption{Relation between absolute $K$ band magnitudes ($M_{K}$) and periods ($P$) for the RSGs in M31. The blue line shows the fitted curve, excluding the 12 points forming a parallel sequence on the left (cf.~Fig.~\ref{fig:mesaPL}, see text). The points are color-coded by their corresponding RMS amplitudes measured from the PTF lightcurves, as shown by the color-bar.}\label{fig:PMk}
\end{figure}

Bound by the limitations of the popular standard methods for our application, we turn to a relatively recent approach based on Gaussian Process (GP) modeling (a non-parametric statistical model that assumes the distribution over a set of random {\em function} values as a multi-variate Gaussian, typically with mean 0; see \citealt{Rasmussen-2005}). In principle, this Bayesian technique provides a statistically robust and powerful way to model any phenomenology, particularly ones for which prior knowledge may not be available.   
For a GP, determining the covariance matrix is the crux of the modeling, which in turn is determined by the power spectrum of the signal. In almost all of the existing off-the-shelf GP-based methods used by the stellar community, a functional form is assumed for generating the covariance matrix elements either assuming stationarity for the lightcurve (i.e., its statistical properties are assumed constant) or a single frequency (for periodic signals, e.g., \citealt{Wang-2012}). Given the complexities of the RSG lightcurves as mentioned above, these algorithms are not accurate for their analysis. 

For our case, the GP modeling is performed by using the so-called critical filter algorithm of signal reconstruction as implemented by \citet{Oppermann-2013} using the \texttt{NIFTy} package of \citet{Selig-2013} aimed toward cosmological signals, and applied to period analysis by Oppermann et al. (2018, to be submitted).
Simply put, the critical filter provides a methodology to reconstruct the underlying signal, whose power spectrum is unknown, by simultaneously also reconstructing this spectrum from the observed data. The computation is iterative. We start with an initial guess of the power spectrum that defines the prior distribution, and reconstruct the signal thus obtaining a posterior distribution conditioned on the observed data and the given power spectrum, which is again a Gaussian. The latter distribution gives an improved estimate of the power spectrum, which is used to update the prior distribution, and then this is used to reconstruct the signal and so on. The iteration is stopped when the new estimate of the power spectrum converges.  Since \citet{Kiss-2006} have illustrated the presence of a 1/$f$ component in the power spectra of these RSGs, this provides a natural choice as our initial guess, but we also impose the condition that there has been at least five iterations before convergence. We obtain a similar reconstruction even when starting with a flat spectrum.  

Examples of the resulting power spectra of the lightcurves are shown in the right panels of Fig.~\ref{fig:lc_spec_1}, with the red-noise component evident. To extract the peaks in a spectrum, we fit a power law to the background red-noise and use that to set a S/N threshold of 7 above the background. We obtain the slopes for these RSGs in the range -1.51 to -0.86. We then select the peaks above this threshold with values greater than $10^{-3}~{\rm day}^{-1}$ (constrained by our baseline). For cases with more than one peak, we choose the one with the maximum power (i.e., the dominant mode) and for a few sources with blended peaks, we take their excess power-weighted average. We adopt these frequencies to define the periods of the lightcurves used in deriving the PL-relation (Sect.~\ref{PL}). 

The methodology we have used reconstructs the lightcurve along the complete time axis, fully taking into account the observational sampling. We have verified by shuffling the magnitude measurements while keeping the same irregular time sampling and calculating the power spectrum of the resulting lightcurve, that no artificial periods corresponding to the observational pattern emerge as power peaks. 

A detailed analysis of how the results from other period-finding algorithms fare against that of the GP modeling is beyond the scope of this paper (it will be presented by Oppermann et al., 2018; to be submitted). Nevertheless, for a simple comparison, we use available \texttt{python} implementations of common period-finding algorithms, specifically LS, supersmoother{\footnote{\url https://www.astroml.org/gatspy/periodic/supersmoother.html}} and PDM, to estimate the periods of some example lightcurves of M31 RSGs. We consider the RSGs with IDs 227 and 189 shown in Fig.~\ref{fig:lc_spec_1}, with quite complicated shapes. The different algorithms give a range of periods: 139~days (LS), 1113~days (supersmoother),  472~days (PDM),  505~days (this work) for RSG\_227;  263~days (LS), 1329~days (supersmoother), 786~days (PDM), 331~days (this work) for RSG\_189. As can be seen from the power spectra of these two RSGs in Fig.~\ref{fig:lc_spec_1}, the results from the other methods, though associated with power peaks in most cases, do not pick up the dominant peaks barring PDM for RSG\_227.

We successfully obtain the {\em pulsation} periods (less than $\approx 1000$~days) for 63 RSGs (the remaining 104 do not have a significant peak), and these are shown in Table~\ref{table_per}. 
We mark these RSGs in Fig.~\ref{fig:Lrms} by the filled red circles, and as is evident, some bright, highly variable sources do not have detectable periods, an example of which is shown in the last panel of Fig.~\ref{fig:lc_spec_1}. It is quite likely that with an extended baseline (for example with the Zwicky Transient Facility, \citealt{Bellm-2014}), a period $>1000$~days could be found for such cases (see also Sect.~\ref{PL}). Another example of a bright, but less variable, non-periodic source is also shown in Fig.~\ref{fig:lc_spec_1}. 
A detailed investigation of why some of the sources in our sample show detectable periods and others, though variable, show none is beyond the scope of this paper. As described above, the mechanism of pulsation in RSGs and its interaction with convection is not well understood.

\begin{figure}[b]
\centering
\includegraphics[width=88mm, height=80mm]{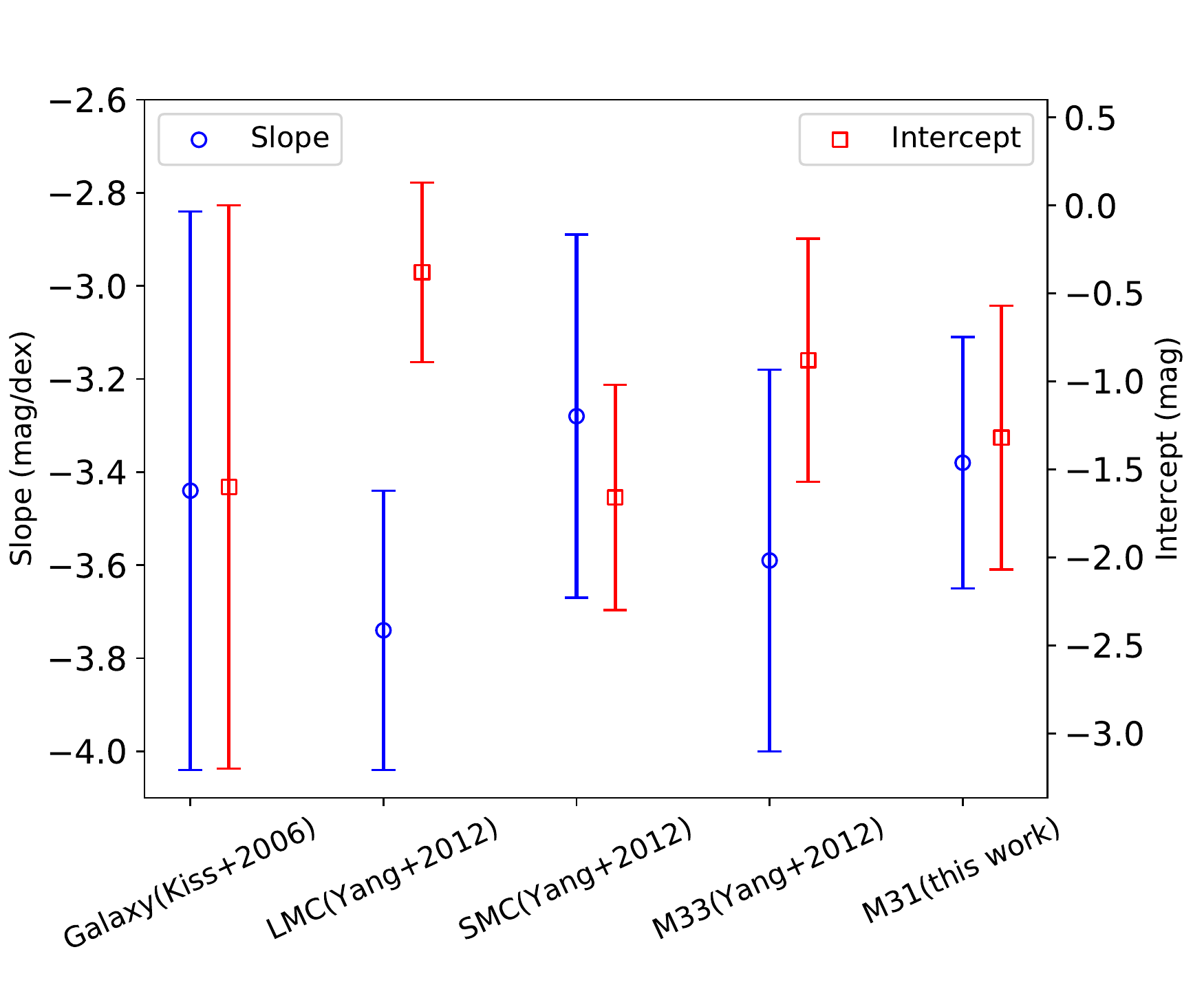}
\caption{Comparison of the slopes (shown in blue) and intercepts (in red) for the $P\mbox{--}M_{K}$ relations of RSGs in different galaxies.}\label{fig:univ}
\end{figure}

\begin{figure*}[t]
\centering
\includegraphics[width=80mm]{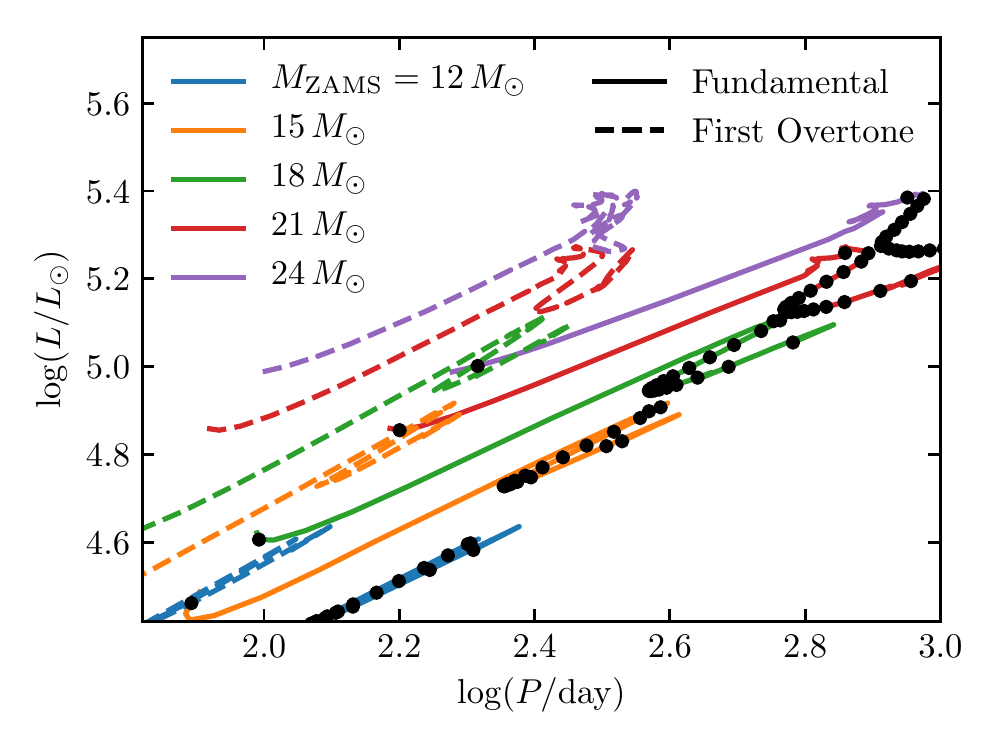}\hfil\includegraphics[width=96mm]{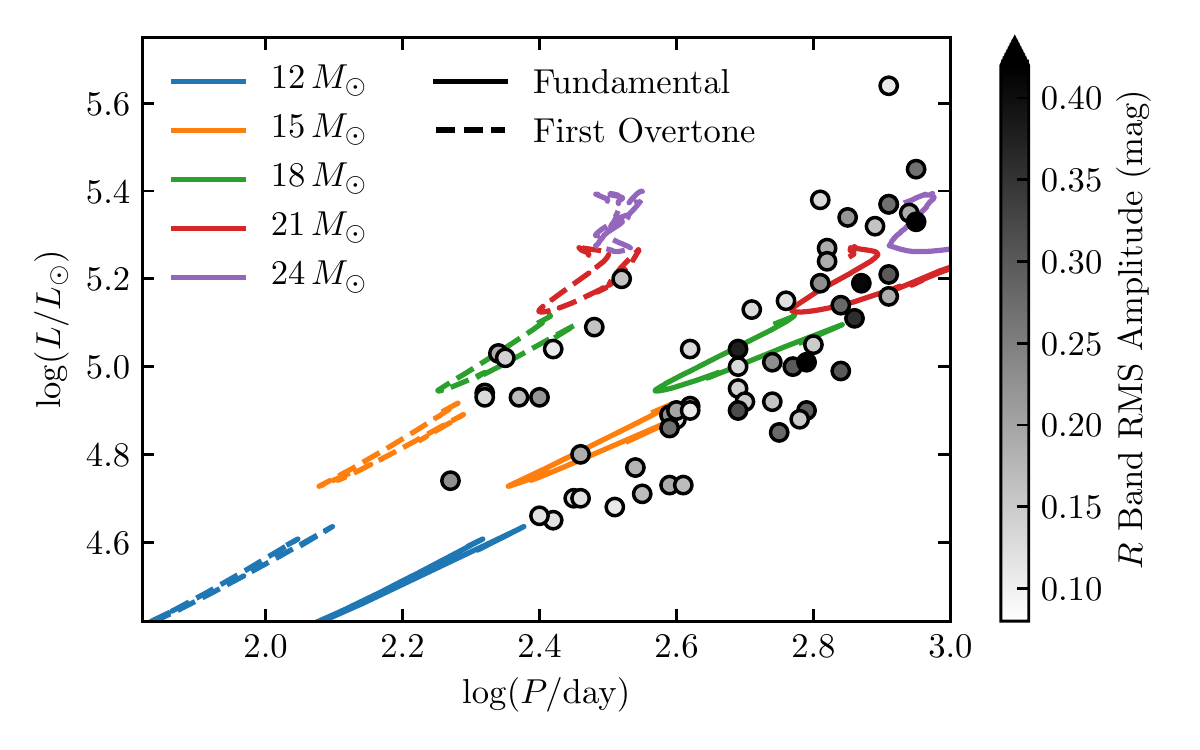}
\caption{{\it Left}: theoretical period-luminosity distribution for RSGs obtained with the \texttt{MESA} models. The black solid points mark equally spaced evolutionary times on the track. Not all 20 points are shown on the track for the lowest mass as some of them are outside the plot boundaries. {\it Right}: same distribution from the left, but with the observed RSGs from Fig.~\ref{fig:PMk} over-plotted as circular points color-coded by their RMS amplitudes  indicated by the greyscale. The theoretical tracks are truncated in this plot as described in the text (Sect.~\ref{mesa}).}\label{fig:mesaPL}
\end{figure*}

\section{Period-luminosity relation of RSGs in M31}\label{PL}

Figure~\ref{fig:PMk} shows the periods of the RSGs obtained in the previous section and their absolute $K$-band magnitudes from ME16 (Sect.~\ref{m31_rsgs}). Of the 63 RSGs with identified periods, 51 appear as a coherent band between $\log(P/{\rm days})\approx2.4$ at M$_{K}\approx-9.5$ and $\log(P/{\rm days})\approx3.0$ at M$_{K}\approx-11.5$. However, 12 RSGs follow a separate sequence (copper circles, leftmost portion of the plot) with 10 of them appearing at similar $M_{K}<-9.5$ as the bulk of the RSGs in the adjacent band, but with an offset of $\log(P/{\rm days})\approx0.3$. The nature of these stars will be examined in Sect.~\ref{mesa}.

The remaining two stars of the 12 have periods $<100$~days (IDs 52 and 166, cf.~Table~\ref{table_per}). Short periods of around 100--150 days have been found for RSGs \citep[e.g.,][]{Kiss-2006}, but no RSGs in the literature have pulsation periods as short as these stars, prompting us to examine them in detail. Both stars possess extremely regular lightcurves as compared to the other RSGs (see Fig.~\ref{fig:lc_spec_1}), relatively high $T_{\rm eff}$ values (4300~K on the temperature scale of ME16), and relatively low luminosities ($M_{K}\ge-9.5$~mag). These physical properties overlap with those observed for ultra-long period Cepheids \citep{Fiorentino2012}. Indeed, star RSG 166 was classified as a Cepheid by \citet{Riess-2012}. As the RSG sample of ME16 was simply selected to have temperature cooler than around 4300~K and be a member of M31 (as opposed to strictly stars on the RSG branch), it is not wholly unexpected that a small number of stars on the luminous and cool edge of the Cepheid instability strip would be identified.

Ignoring the 12 RSGs from above, we obtain a simple linear fit through the rest as 
\begin{equation}
M_{K}=(-3.38\pm0.27)\times \log P+(-1.32\pm0.75),
\end{equation}
shown by the blue line in Fig.~\ref{fig:PMk}. 
The RSG $P\mbox{--}M_{K}$ relation we have derived here is the first for M31. We see a dispersion around this relation of 0.29~mag. A part of it is possibly due to the errors in the adopted $K$ band magnitudes, which could reach a tenth of a magnitude (for example, the 2MASS values given by \citealt{Massey-2009} and used in ME16), and the fact that these $K$ band magnitudes are from single (nightly) observation epochs, while the amplitude in this band could be as high as 0.25~mag \citep{Wood-1983} (thus, a contribution of $<0.18$~mag to the dispersion). Also, circumstellar reddening may contribute to this dispersion; ME16, however, applied an average correction for M31 RSGs obtained by \citet{Massey-2009} to the $M_{K}$ values of these stars that should reduce this effect. But a possible part of the cause could also be that the dispersion is intrinsic for these objects---an apparent effect of additional parameters (cf.~Sect.~\ref{mesa}).

Given the broad trend of increasing variability with brightness (Fig.~\ref{fig:Lrms}) discussed in Sect.~\ref{lcs}, and the positive correlation of period with brightness (Fig.~\ref{fig:PMk}), the same broad trend can be expected between the periods and the RMS deviations. This is indicated in Fig.~\ref{fig:PMk} by the gradient in the face-color of the points. In light of such a trend, it is feasible that a period longer than the range probed in this study is present in the highly variable sources without detectable periods (Fig.~\ref{fig:lc_spec_1}, Sect.~\ref{period}). Furthermore, the theoretical result of increasing pulsation amplitude with initial masses or luminosities \citep[e.g.,][]{Yoon-2010} appears consistent with our observational result.

Figure~\ref{fig:univ} shows the comparison of our results for the $P\mbox{--}M_{K}$ relation with those of others for different galaxies. Our slope and intercept are consistent with those of \cite{Kiss-2006} for the Galactic RSG sample with 13 sources and a dispersion of 0.46~mag. Furthermore, they are in good agreement with those of the LMC, SMC, and M33 derived by \citet{Yang-2012} using 47, 21 and 40 sources, respectively, with dispersion value of $\approx 0.3$~mag (they claim their relation for the Galactic sample is less accurate due to uncertainties in distances)\footnote{\citet{Yang-2012} adopted distance moduli of 18.50 (LMC), 18.91 (SMC) and 24.93 (M33).}. Thus, the data from several galaxies characterized by different metallicities (from sub-solar by a factor of $\approx 4$ in the Small Magellanic Cloud to super-solar by a factor of $\approx 2$ in M31, e.g., \citealt{Massey-2003}), are all consistent with a universal period-luminosity relation for RSGs, although the uncertainties still allow a wide range of parameters. Such universality may not be unexpected given current ideas on what drives the pulsation in RSGs ($\kappa$ mechanism in hydrogen, Sect.~\ref{mesa}), and this potentially could have important implications for the life of very low-metallicity/Population III stars \citep[e.g.,][]{Moriya-2015}.

\section{\texttt{MESA} models: physical parameters of RSGs}\label{mesa}
We investigate the pulsation mode represented by the periods we have found for the RSGs with theoretical models constructed using Modules for Experiments in Stellar Astrophysics (\texttt{MESA}; \citealt{Mesa-1,Mesa-2, Mesa-3, Mesa-4}), a standard tool for stellar evolutionary calculations. One of the latest updates presented by \citet{Mesa-3} includes fully incorporating existing methodology for asteroseismology computations (\texttt{GYRE} software of \citealt{Townsend-2013}), and thus facilitates comparison of our observational results with theory.

To produce theoretical predictions for RSG pulsation periods, we use version 9793 of the \texttt{MESA} software. We consider stellar models with ZAMS masses of $12\mbox{--}24~M_{\odot}$, which are evolved from ZAMS to depletion of Carbon in the core. Our \texttt{MESA} settings follow the recent work of \citet{Chun17}, including a calibrated mixing length parameter $\alpha_{\rm MLT} = 2.7$ to match inferred RSG temperatures in M31 at $Z = 0.04$. We also include step overshoot with parameter $f_{\rm ov} = 0.15$ and mass loss according to the ``Dutch'' wind prescription \citep{deJager88,Vink01}. More details on \texttt{MESA} settings can be found in \citet{Chun17}. Our models do not include any rotation.

After the models leave the main sequence, we include adiabatic \texttt{GYRE} calculations in each step for $T_{\rm eff}<4500$~K. These calculations find all radial modes with periods in the range 10--10000 days, more than adequate to identify the fundamental mode and first overtone in these RSG models. The left panel of Fig.~\ref{fig:mesaPL} shows the resulting period tracks for several different mass models, along with points representing 20 equally spaced time intervals along each of the tracks for the fundamental mode periods. When comparing to the observed RSG data in the right panel of the same figure, we show the same tracks only after the onset of core helium burning, where the models spend more than 95\% of their RSG lifetime.

Our modeling makes no attempt to discern which modes should be excited. The work of \cite{Heger-1997} suggests that RSG pulsations are excited by the $\kappa$ mechanism in the hydrogen
ionization zone. \cite{Heger-1997} also find that pulsation periods roughly scale as ${P \propto L/M}$ in agreement with the analytic predictions of \cite{Gough-1965}. Our \texttt{MESA} models roughly agree with this scaling.

The right panel of Fig.~\ref{fig:mesaPL} shows that the bulk of the pulsating RSGs in our sample are consistent with the fundamental mode pulsation of the theoretical models. Furthermore, the sequence of 10 stars offset to shorter periods for a given $M_{K}$, is broadly consistent with the first overtone mode.

\begin{figure}
\centering
\includegraphics[width=90mm]{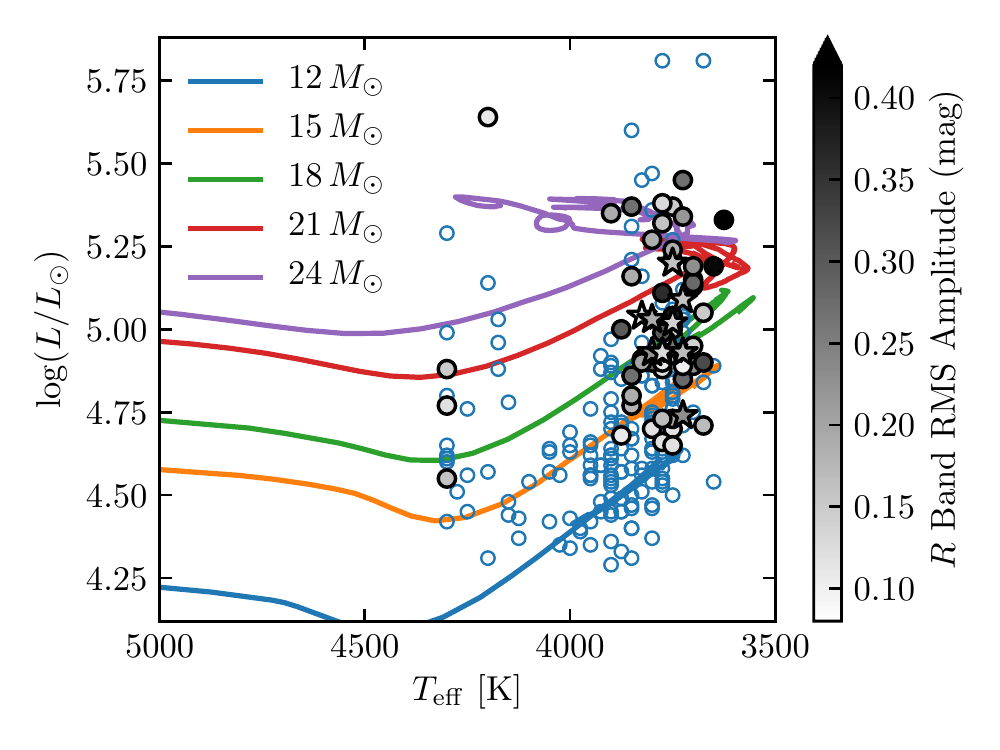}
\caption{HR diagram for the \texttt{MESA} evolutionary tracks and the RSGs with detected periods, shown here as grey points. The star symbols denote the sequence of RSGs offset to shorter periods (Fig.~\ref{fig:mesaPL}).    
Unfilled blue circles show objects for which significant periods have not been identified.}\label{fig:hrd}
\end{figure}

The Hertzsprung-Russell (HR) diagram for cool supergiants is shown in Fig.~\ref{fig:hrd}. Colored lines represent the same \texttt{MESA} models used to produce Fig.~\ref{fig:mesaPL}, and the RSGs with measured pulsation periods are shown in grey (other M31 RSGs are plotted as blue circles for reference). The observed RSGs lie in proximity to \texttt{MESA} model tracks for broadly the same initial masses, whether period or $T_{\rm eff}$ is used as parameter (right panel of Fig.~\ref{fig:mesaPL} and Fig.~\ref{fig:hrd}).  
The 10 stars with measured periods offset from the bulk of the sample in the period-luminosity plane that are not Cepheids are shown by the star symbols in the HR diagram. These stars with shorter periods overlap with the bulk of the RSG population with measured periods in their temperatures and luminosities. This strengthens our association of the variability observed in these stars with the first overtone mode of stars with the same properties as the bulk population.

As is already known, there is no single-valued mass-luminosity relation for the red supergiants. Consequently, tracks of different masses at the same luminosity can smear out the period for the RSGs, and may be a reason for the scatter seen in Fig.~\ref{fig:PMk}. Thus, the mass of the star, and more importantly its evolutionary stage, naturally appear as additional parameters contributing to some of the dispersion in the $P-M_{K}$ relation. As a result, the precision of any extragalactic distance estimate based on this relation can never be better than the limit imposed by the intrinsic dispersion.

\section{Effect of intrinsic variability on SN~II-P progenitor mass estimates}\label{sne_mass}

RSGs as progenitors of Type II-P/L SNe \citep[e.g.,][]{Ekstrom-2012} have been established based, in part, on the direct detection of some progenitor stars in archival images of nearby galaxies hosting these SNe \citep[e.g.,][]{Smartt-1, Van-Dyk-2003}. A recent comprehensive review of such observations spanning a 15 year period, and their implications for the progenitor masses of hydrogen-rich core-collapse supernovae, was presented by \citet{Smartt-2} [hereafter SS15]. The general approach to determining progenitor masses from these pre-explosion observations is to convert the photometric measurements to bolometric luminosities and then compare to the initial mass-final luminosity relation from stellar models. One of the important results from such an analysis is the maximum-mass for these progenitors is claimed to be at a much lower value ($\approx 18~M_{\odot}$)  (\citealt{Smartt-2009}; SS15) than the high-mass limit of RSGs ($25\mbox{--}30~M_{\odot}$) predicted by stellar evolution theory. 

However, various systematic errors, often resulting from the uncertain final stages in RSG evolution, can influence this result. In particular, when converting observed fluxes in a limited (often single) photometric band to bolometric luminosities, treatments of extinction and bolometric corrections are critical. For example, \citet{Davies-2017} showed that the spectral type of RSGs evolves to later type as they approach core collapse, resulting in a larger bolometric correction than typically assumed for the imaged progenitor stars. They thus pointed out that failing to account for the spectral evolution of RSGs can introduce a systematic error in the adopted bolometric correction.  

Most of the archival detections of supernova progenitors are from single-epoch observations, 
and we have demonstrated in Sect.~\ref{lcs} (see Fig.~\ref{fig:Lrms}) that while lower luminosity RSGs have observed variability of $\lesssim (5\mbox{--}10)\%$, a large fraction of the RSGs with luminosities above $\log(L/L_{\odot})\approx4.8$ show a significant variability. Thus, if our observed sample of M31 RSGs is representative, intrinsic variability of RSGs will {\em preferentially} affect the initial mass estimates of high luminosity (and therefore high initial mass) progenitor stars. It is therefore pertinent to examine how big such an effect can be.

Recently \citet{Johnson-2017} (see also \citealt{Kochanek-2017}) examined the variability of four SN~II-P progenitor stars, from approximately four to eight years prior to the supernovae. They derived upper limits on their variability at (5--10)\% of their $R$ band luminosities, corresponding to $\Delta m_{R}<0.05\mbox{--}0.10$. Three of these SN progenitors had sufficient observations to derive bolometric luminosities, all of which were in the range $\log(L/L_{\odot})=4.5-4.9$ (\citealt{Davies-2017, Kochanek-2017, van-Dyk-2017}). 
Thus, as is evident from Fig.~\ref{fig:Lrms} (right panel), the upper limits on their variability are also consistent with our results for the RSGs in M31.

\begin{figure}[t]
\centering
\includegraphics[width=88mm]{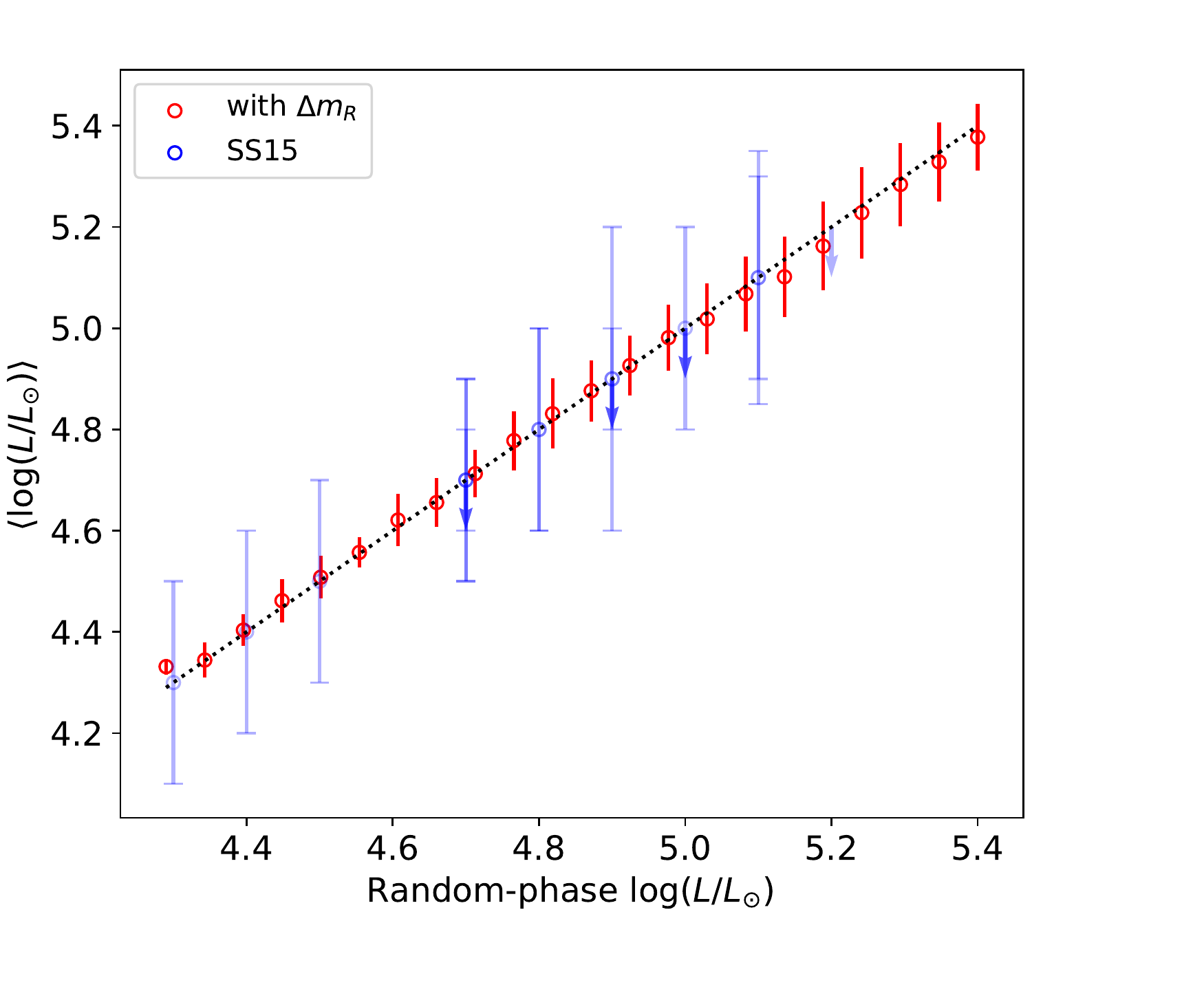}
\caption{Mean bolometric luminosity $\left<\log (L/L_{\odot})\right>$ vs. random-phase bolometric luminosity of RSGs in M31 (shown in red). The vertical bars for these red points indicate the {\em standard deviations} of the distributions of $\left<\log (L/L_{\odot})\right>$ at the given random-phase luminosity. The black dotted line shows the 1:1 relation between the two parameters, i.e., assuming a random-phase luminosity as the mean value---the relevant scenario when using single epoch archival image measurements of progenitor stars, and the blue points denote this for the SN~II-P progenitors from SS15, with the vertical bars in this case denoting {\em measurement errors}.}\label{fig:rand_mag}
\end{figure}

In this section, we make a quantitative assessment of whether intrinsic variability of RSGs can have any significant effect on the SN progenitor mass estimates.  

In converting a photometric measurement in some filter ($m_{\lambda}$) to the bolometric luminosity ($\mathcal{L}=\log (L/L_{\odot})$), a correction (${\rm d}m_{\lambda}$ to obtain the mean value $\left<m_{\lambda}\right>$) for possible variability of the progenitor can be applied as
\begin{equation}
2.5\cdot\left<\mathcal{L}\right>=M_{\odot}-\left(m_{\lambda}+{\rm d}m_{\lambda}-A_{\lambda}+{\rm BC}_{\lambda}\right),
\label{bol}
\end{equation}
where the term inside the brackets in the right-hand side is the bolometric magnitude of the star, $A_{\lambda}$ and ${\rm BC}_{\lambda}$ are the extinction and bolometric corrections, respectively, for the star in the given filter, and  $M_{\odot}$ is the bolometric magnitude of the sun. In absence of color information to constrain the ${\rm BC}_{\lambda}$ of the star, typically the value for an M0 supergiant is assumed with its $1\sigma$ error as 0.3~mag corresponding to the standard deviation of the bolometric corrections spanning the spectral types from late K to late M (see \citealt{Smartt-2009}, SS15).

The detections and upper limits of the progenitor stars have been acquired mostly in the $I$ and $R$ bands. Since we do not have information on the intrinsic variability of our RSG sample in the $I$ band, we assume a similar level of  variability as in the neighboring $R$ band (Fig.~\ref{fig:Lrms}). Then, given a random-phase measurement $m_{\lambda}$ from the pre-explosion image of the star, and hence, a random-phase luminosity $\mathcal{L}$ computed from it, we are interested in the mean luminosity $\left<\mathcal{L}\right>$ for the star. To this end, we use Bayes' Theorem to derive the {\em posterior probability distribution} for the mean luminosity $\left<\mathcal{L}\right>$, given the observed random-phase luminosity $\mathcal{L}$, according to
\begin{equation}
P(\left<\mathcal{L}\right>|\mathcal{L})=\frac{P(\mathcal{L}|\left<\mathcal{L}\right>)\cdot P(\left<\mathcal{L}\right>)}{P(\mathcal{L})}.
\label{Bayes}
\end{equation}

The luminosity distribution of our RSG sample is taken as the prior $P(\left<\mathcal{L}\right>)$. We obtain a similar result when using a flat prior. To construct the likelihood $P(\mathcal{L}|\left<\mathcal{L}\right>)$, we use our full sample of M31 RSGs with luminosity available from ME16, i.e., all the RSGs in Fig.~\ref{fig:static} both below and above the PTF variability sensitivity limit. We bin the stars in $\log L$  with bin-width $\approx 0.1$ (Fig.~\ref{fig:Lrms}), and compute the distribution of ${\rm d}m=m-\left<m\right>$ magnitudes using the lightcurves of all the stars in each bin. Our construction of {\em relative fluxes} in the PTF $R$ band assumed these relative fluxes are similar in the other filters where most progenitor images were taken. 
We scale the resulting distribution in the corresponding bin containing $\left<\mathcal{L}\right>$ by a factor 2.5 and shift it by $\left<\mathcal{L}\right>$ (cf.~Eq.~\ref{bol}) to obtain the likelihood. The denominator in Eq.~\ref{Bayes} is just a normalization constant. With all these ingredients, we compute the posterior distribution $P(\left<\mathcal{L}\right>|\mathcal{L})$ and then obtain the mean and standard deviation of this distribution.

The result is shown in Fig.~\ref{fig:rand_mag} by the red points. The diagonal dotted line in the figure shows the 1:1 relation between the random-phase $\mathcal{L}$ and the mean $\left<\mathcal{L}\right>$, and thus represents the typical assumption made when computing supernova progenitor masses from pre-explosion images.

From this figure, we see that $\left<\mathcal{L}\right>$ vs. $\mathcal{L}$ for our RSG sample is consistent with the 1:1 relation. 
The vertical bars of the red points represent standard deviations of the $\left<\mathcal{L}\right>$ distributions, and as can be seen, they become larger with increasing luminosity, reflecting the fact that the observed variability in our sample is larger for higher luminosity stars (Fig.~\ref{fig:Lrms}). We also plot the measured luminosities along with the {\em measurement errors} of the SN~II-P progenitor stars given by SS15 with blue points based on the 1:1 relation. As is evident, the measurement errors (from the photometry, extinction and bolometric corrections, cf.~Eq.~\ref{bol}) of the random-phase luminosities of the progenitors are much larger than the standard deviations of the $\left<\mathcal{L}\right>$ distributions.
The overall shift in the masses of the progenitors accounting for the intrinsic variability (a mere $1\mbox{--}2~M_{\odot}$) is not significant, given the rather large uncertainty on the applied corrections ($\Delta A_{\lambda}$ and $\Delta {\rm BC}_{\lambda}$) and photometric measurements ($\Delta m_{\lambda}$). The uncertainty due to variability (${\rm d}m$), however, is an inevitable physical effect that cannot be overcome by taking more precise measurements and thus will eventually become the dominant effect.

The analysis presented above rests on the assumption that our full sample of RSGs in M31 is representative of the variability that RSGs exhibit in the final years before core-collapse. If these properties change as a RSG enter the final nuclear burning stages, our results and conclusions could be influenced. Further, we have assumed that the variability of RSGs in the PTF $R$ band is similar in the filters (largely $I$) that the SN~II-P progenitors were imaged. If RSGs were markedly more variable in the $I$ band, then our results could be affected. Nevertheless, there is no obvious argument for variability amplitude to be larger in $I$ than in $R$.

While intrinsic variability of RSGs with respect to luminosity estimates of SN~II-P progenitors is shown here to be not significant, it could, however, play an important role in other physical aspects of RSGs of consequence to their initial mass estimates, in particular a possible enhancement of the mass loss \citep{Heger-1997, Yoon-2010}. A similar effect is known to take place in Miras \citep[e.g.,][]{Bowen-1988} and in Cepheids \citep{Neilson-2008}, wherein variability due to radial pulsations and the associated lifting and lowering of extended atmospheric layers is regarded to be an important driver for mass loss.  
Mutli-epoch monitoring, in multiple filters, for variability of potential SN~II-P progenitors, like the efforts of \citet{Gerke-2015}, \citet{Kochanek-2017}, \citet{Johnson-2017} will continue to be beneficial and have implications beyond the photometric uncertainty of the progenitor stars.

\section{Conclusions}\label{conclude}
The Palomar Transient Factory's nearly 2000 days of monitoring of M31 has allowed for the first measurements of the prevalence of RSG variability in this important nearby galaxy. In this first study, we confined our efforts to the 255 bright RSGs that were cataloged and typed by the LGGS group (ME16). As was known for our galaxy, the
Magellanic Clouds and M33, these evolved massive stars can be highly variable, and often periodic. Based on the RMS deviations of the PTF $R$ band lightcurves, 167 of the 255 RSGs exhibit variability above the noise limit of PTF data (Fig.~\ref{fig:static}). For these sources, we see a broad positive correlation between the variability amplitudes, represented by the RMS values, and their brightness.

We found that the lightcurves of these variables are characterized by semi-regular (and irregular for some cases) long-period pulsations over a strong red noise \citep{Kiss-2006}.  Using a robust, probabilistically motivated methodology for reconstruction of their power spectra, we determined significant pulsational periods for 63 of the variable RSGs.  Using the absolute magnitudes measured in the $K$ band by ME16, we derived the $P\mbox{--}M_{K}$ relation for the RSGs in M31 and showed that it is consistent with that of the Galaxy, the Magellanic Clouds, and the M33 RSG samples, which represent a wide range of metallicities. This points toward a universality of this relation, at least for the Local Universe. It will be interesting to explore this relation with a large sample of RSGs from a larger number of distant resolved galaxies with the upcoming large telescopes like the Large Synoptic Survey Telescope (LSST; \citealt{Ivezic-2008}) and the European Extremely Large Telescope \citep{Greggio-2012}. 

Comparison of the measured Period Luminosity relation with the theoretical one based on the \texttt{MESA} models shows that the pulsations are consistent with the fundamental radial mode. At the same time, a clear sequence of outliers is seen in the distribution with our observed sample, which is found to be consistent with the first overtone mode based on the models---another novel feat of this work.

Finally, we explored whether the measured RSG variability could be large enough to confound the inference of stellar mass from the observed supernova progenitor luminosity. We found that this effect is small compared to current observational uncertainties. 
 
Future insights on RSG variability in M31 with the existing PTF data could be gained from large-scale variability studies of bright stars not necessarily in the ME16 catalog. In addition, with the recent commissioning of the ZTF instrument, continued monitoring of all RSGs will be possible and hopefully reveal longer period pulsations than possible with the original data set. Moreover, variability observations of different RSG populations spanning near and far host galaxies will be an asset for conducting many important studies of these stars. LSST in the near future, with its superb photometry and expected cadence of three days, has the potential to generate such a product from monitoring a select group of galaxies.

\acknowledgments
{\it Acknowledgments}: This work was supported by the National Science Foundation through grants PHY 11-25915, PHY 17-148958 and ACI 16-63688 and funded in part by the Gordon and Betty Moore Foundation through Grant GBMF5076.  
MS is grateful to Niels Oppermann for the detailed discussions on statistical approaches and methods of proper signal reconstruction. MS also thanks Mathew Graham and Ashish Mahabal for helpful discussions on the period finding algorithms, and Matteo Cantiello for the helpful suggestions and comments on the manuscript. EB thanks Sang-Hyun Chun for providing \texttt{MESA} inlists necessary to reproduce their RSG models, and Matteo Cantiello for discussion of RSG evolution in \texttt{MESA}. PyFITS is a product of the Space Telescope Science Institute, which is operated by AURA for NASA. Support for this work was provided to MRD by NASA through Hubble Fellowship grant NSG-HF2-51373 awarded by the Space Telescope Science Institute, which is operated by the Association of Universities for Research in Astronomy, Inc., for NASA, under contract NAS5-26555. MRD acknowledges support from the Dunlap Institute at the University of Toronto.

\software{\texttt{numpy} \citep{numpy}, \texttt{scipy} \citep{scipy}, \texttt{astropy} \citep{astropy}, \texttt{pyfits}, \texttt{pandas} \citep{pandas}, \texttt{mpi4py} \citep{mpi4py}, \texttt{matplotlib} \citep{matplotlib}, \texttt{DAOPHOT} \citep{Stetson}, \texttt{DAOGROW} \citep{Stetson-1990},\texttt{NIFTy} \citep{Selig-2013}.}

\bibliographystyle{yahapj}
\bibliography{references}

\begin{thebibliography}{}
\providecommand\natexlab[1]{#1}
\providecommand\JournalTitle[1]{#1}

\bibitem[{{Alcock} {et~al.}(1997){Alcock}, {Allsman}, {Alves}, {Axelrod},
  {Becker}, {Bennett}, {Cook}, {Freeman}, {Griest}, {Guern}, {Lehner},
  {Marshall}, {Peterson}, {Pratt}, {Quinn}, {Rodgers}, {Stubbs}, {Sutherland},
  \& {Welch}}]{Alcock-1997}
{Alcock}, C., {Allsman}, R.~A., {Alves}, D., {et~al.} 1997,
  \href{http://dx.doi.org/10.1086/304535}{\JournalTitle{\apj}, 486, 697}

\bibitem[{{Arnett} {et~al.}(2014){Arnett}, {Meakin}, \&
  {Viallet}}]{Arnett-2014}
{Arnett}, W.~D., {Meakin}, C., \& {Viallet}, M. 2014,
  \href{http://dx.doi.org/10.1063/1.4867384}{\JournalTitle{AIP Advances}, 4,
  041010}

\bibitem[{{Astropy Collaboration} {et~al.}(2013){Astropy Collaboration},
  {Robitaille}, {Tollerud}, {Greenfield}, {Droettboom}, {Bray}, {Aldcroft},
  {Davis}, {Ginsburg}, {Price-Whelan}, {Kerzendorf}, {Conley}, {Crighton},
  {Barbary}, {Muna}, {Ferguson}, {Grollier}, {Parikh}, {Nair}, {Unther},
  {Deil}, {Woillez}, {Conseil}, {Kramer}, {Turner}, {Singer}, {Fox}, {Weaver},
  {Zabalza}, {Edwards}, {Azalee Bostroem}, {Burke}, {Casey}, {Crawford},
  {Dencheva}, {Ely}, {Jenness}, {Labrie}, {Lim}, {Pierfederici}, {Pontzen},
  {Ptak}, {Refsdal}, {Servillat}, \& {Streicher}}]{astropy}
{Astropy Collaboration}, {Robitaille}, T.~P., {Tollerud}, E.~J., {et~al.} 2013,
  \href{http://dx.doi.org/10.1051/0004-6361/201322068}{\JournalTitle{\aap},
  558, A33}

\bibitem[{{Beasor} \& {Davies}(2016)}]{Beasor-2016}
{Beasor}, E.~R., \& {Davies}, B. 2016,
  \href{http://dx.doi.org/10.1093/mnras/stw2054}{\JournalTitle{\mnras}, 463,
  1269}

\bibitem[{{Beasor} \& {Davies}(2017)}]{Beasor-2017}
---. 2017, \JournalTitle{ArXiv e-prints},
  \href{http://arxiv.org/abs/1712.01852}{{\sffamily arXiv:1712.01852
  [astro-ph.SR]}}

\bibitem[{{Bellm}(2014)}]{Bellm-2014}
{Bellm}, E. 2014, in The Third Hot-wiring the Transient Universe Workshop, ed.
  P.~R. {Wozniak}, M.~J. {Graham}, A.~A. {Mahabal}, \& R.~{Seaman}, 27

\bibitem[{{Bowen}(1988)}]{Bowen-1988}
{Bowen}, G.~H. 1988,
  \href{http://dx.doi.org/10.1086/166378}{\JournalTitle{\apj}, 329, 299}

\bibitem[{{Chun} {et~al.}(2018){Chun}, {Yoon}, {Jung}, {Kim}, \&
  {Kim}}]{Chun17}
{Chun}, S.-H., {Yoon}, S.-C., {Jung}, M.-K., {Kim}, D.~U., \& {Kim}, J. 2018,
  \href{http://dx.doi.org/10.3847/1538-4357/aa9a37}{\JournalTitle{\apj}, 853,
  79}

\bibitem[{{Cutri} {et~al.}(2003){Cutri}, {Skrutskie}, {van Dyk}, {Beichman},
  {Carpenter}, {Chester}, {Cambresy}, {Evans}, {Fowler}, {Gizis}, {Howard},
  {Huchra}, {Jarrett}, {Kopan}, {Kirkpatrick}, {Light}, {Marsh}, {McCallon},
  {Schneider}, {Stiening}, {Sykes}, {Weinberg}, {Wheaton}, {Wheelock}, \&
  {Zacarias}}]{Cutri-2003}
{Cutri}, R.~M., {Skrutskie}, M.~F., {van Dyk}, S., {et~al.} 2003,
  \JournalTitle{VizieR Online Data Catalog}, 2246

\bibitem[{Dalc\'in {et~al.}(2005)Dalc\'in, Paz, \& Storti}]{mpi4py}
Dalc\'in, L., Paz, R., \& Storti, M. 2005,
  \href{http://dx.doi.org/https://doi.org/10.1016/j.jpdc.2005.03.010}{\JournalTitle{Journal
  of Parallel and Distributed Computing}, 65, 1108 }

\bibitem[{{Davies} \& {Beasor}(2017)}]{Davies-2017}
{Davies}, B., \& {Beasor}, E. 2017, \JournalTitle{ArXiv e-prints},
  \href{http://arxiv.org/abs/1709.06116}{{\sffamily arXiv:1709.06116
  [astro-ph.SR]}}

\bibitem[{{Davies} {et~al.}(2013){Davies}, {Kudritzki}, {Plez}, {Trager},
  {Lan{\c c}on}, {Gazak}, {Bergemann}, {Evans}, \& {Chiavassa}}]{Davies-2013}
{Davies}, B., {Kudritzki}, R.-P., {Plez}, B., {et~al.} 2013,
  \href{http://dx.doi.org/10.1088/0004-637X/767/1/3}{\JournalTitle{\apj}, 767,
  3}

\bibitem[{{de Jager} {et~al.}(1988){de Jager}, {Nieuwenhuijzen}, \& {van der
  Hucht}}]{deJager88}
{de Jager}, C., {Nieuwenhuijzen}, H., \& {van der Hucht}, K.~A. 1988,
  \JournalTitle{\aaps}, 72, 259

\bibitem[{{Ekstr{\"o}m} {et~al.}(2012){Ekstr{\"o}m}, {Georgy}, {Eggenberger},
  {Meynet}, {Mowlavi}, {Wyttenbach}, {Granada}, {Decressin}, {Hirschi},
  {Frischknecht}, {Charbonnel}, \& {Maeder}}]{Ekstrom-2012}
{Ekstr{\"o}m}, S., {Georgy}, C., {Eggenberger}, P., {et~al.} 2012,
  \href{http://dx.doi.org/10.1051/0004-6361/201117751}{\JournalTitle{\aap},
  537, A146}

\bibitem[{{Feast} {et~al.}(1980){Feast}, {Catchpole}, {Carter}, \&
  {Roberts}}]{Feast-1980}
{Feast}, M.~W., {Catchpole}, R.~M., {Carter}, B.~S., \& {Roberts}, G. 1980,
  \href{http://dx.doi.org/10.1093/mnras/193.2.377}{\JournalTitle{\mnras}, 193,
  377}

\bibitem[{{Fiorentino} {et~al.}(2012){Fiorentino}, {Clementini}, {Marconi},
  {Musella}, {Saha}, {Tosi}, {Contreras Ramos}, {Annibali}, {Aloisi}, \& {van
  der Marel}}]{Fiorentino2012}
{Fiorentino}, G., {Clementini}, G., {Marconi}, M., {et~al.} 2012,
  \href{http://dx.doi.org/10.1007/s10509-012-1043-4}{\JournalTitle{\apss}, 341,
  143}

\bibitem[{{Fliri} {et~al.}(2006){Fliri}, {Riffeser}, {Seitz}, \&
  {Bender}}]{Fliri-2006}
{Fliri}, J., {Riffeser}, A., {Seitz}, S., \& {Bender}, R. 2006,
  \href{http://dx.doi.org/10.1051/0004-6361:20042223}{\JournalTitle{\aap}, 445,
  423}

\bibitem[{{Georgy}(2012)}]{Georgy-2012}
{Georgy}, C. 2012,
  \href{http://dx.doi.org/10.1051/0004-6361/201118372}{\JournalTitle{\aap},
  538, L8}

\bibitem[{{Gerke} {et~al.}(2015){Gerke}, {Kochanek}, \& {Stanek}}]{Gerke-2015}
{Gerke}, J.~R., {Kochanek}, C.~S., \& {Stanek}, K.~Z. 2015,
  \href{http://dx.doi.org/10.1093/mnras/stv776}{\JournalTitle{\mnras}, 450,
  3289}

\bibitem[{{Glass}(1979)}]{Glass-1979}
{Glass}, I.~S. 1979,
  \href{http://dx.doi.org/10.1093/mnras/186.2.317}{\JournalTitle{\mnras}, 186,
  317}

\bibitem[{{Gough} {et~al.}(1965){Gough}, {Ostriker}, \& {Stobie}}]{Gough-1965}
{Gough}, D.~O., {Ostriker}, J.~P., \& {Stobie}, R.~S. 1965,
  \href{http://dx.doi.org/10.1086/148450}{\JournalTitle{\apj}, 142, 1649}

\bibitem[{{Greggio} {et~al.}(2012){Greggio}, {Falomo}, {Zaggia}, {Fantinel}, \&
  {Uslenghi}}]{Greggio-2012}
{Greggio}, L., {Falomo}, R., {Zaggia}, S., {Fantinel}, D., \& {Uslenghi}, M.
  2012, \href{http://dx.doi.org/10.1086/667195}{\JournalTitle{\pasp}, 124, 653}

\bibitem[{{Groh} {et~al.}(2013){Groh}, {Meynet}, {Georgy}, \&
  {Ekstr{\"o}m}}]{Groh-2013}
{Groh}, J.~H., {Meynet}, G., {Georgy}, C., \& {Ekstr{\"o}m}, S. 2013,
  \href{http://dx.doi.org/10.1051/0004-6361/201321906}{\JournalTitle{\aap},
  558, A131}

\bibitem[{{Guo} \& {Li}(2002)}]{Guo-2002}
{Guo}, J.~H., \& {Li}, Y. 2002,
  \href{http://dx.doi.org/10.1086/324295}{\JournalTitle{\apj}, 565, 559}

\bibitem[{{Heger} {et~al.}(2003){Heger}, {Fryer}, {Woosley}, {Langer}, \&
  {Hartmann}}]{Heger-2003}
{Heger}, A., {Fryer}, C.~L., {Woosley}, S.~E., {Langer}, N., \& {Hartmann},
  D.~H. 2003, \href{http://dx.doi.org/10.1086/375341}{\JournalTitle{\apj}, 591,
  288}

\bibitem[{{Heger} {et~al.}(1997){Heger}, {Jeannin}, {Langer}, \&
  {Baraffe}}]{Heger-1997}
{Heger}, A., {Jeannin}, L., {Langer}, N., \& {Baraffe}, I. 1997,
  \JournalTitle{\aap}, 327, 224

\bibitem[{Hunter(2007)}]{matplotlib}
Hunter, J.~D. 2007,
  \href{http://dx.doi.org/10.1109/MCSE.2007.55}{\JournalTitle{Computing In
  Science \& Engineering}, 9, 90}

\bibitem[{{Ivezic} {et~al.}(2008){Ivezic}, {Tyson}, {Abel}, {Acosta},
  {Allsman}, {AlSayyad}, {Anderson}, {Andrew}, {Angel}, {Angeli}, {Ansari},
  {Antilogus}, {Arndt}, {Astier}, {Aubourg}, {Axelrod}, {Bard}, {Barr},
  {Barrau}, {Bartlett}, {Bauman}, {Beaumont}, {Becker}, {Becla}, {Beldica},
  {Bellavia}, {Blanc}, {Blandford}, {Bloom}, {Bogart}, {Borne}, {Bosch},
  {Boutigny}, {Brandt}, {Brown}, {Bullock}, {Burchat}, {Burke}, {Cagnoli},
  {Calabrese}, {Chandrasekharan}, {Chesley}, {Cheu}, {Chiang}, {Claver},
  {Connolly}, {Cook}, {Cooray}, {Covey}, {Cribbs}, {Cui}, {Cutri}, {Daubard},
  {Daues}, {Delgado}, {Digel}, {Doherty}, {Dubois}, {Dubois-Felsmann},
  {Durech}, {Eracleous}, {Ferguson}, {Frank}, {Freemon}, {Gangler}, {Gawiser},
  {Geary}, {Gee}, {Geha}, {Gibson}, {Gilmore}, {Glanzman}, {Goodenow},
  {Gressler}, {Gris}, {Guyonnet}, {Hascall}, {Haupt}, {Hernandez}, {Hogan},
  {Huang}, {Huffer}, {Innes}, {Jacoby}, {Jain}, {Jee}, {Jernigan},
  {Jevremovic}, {Johns}, {Jones}, {Juramy-Gilles}, {Juric}, {Kahn}, {Kalirai},
  {Kallivayalil}, {Kalmbach}, {Kantor}, {Kasliwal}, {Kessler}, {Kirkby},
  {Knox}, {Kotov}, {Krabbendam}, {Krughoff}, {Kubanek}, {Kuczewski},
  {Kulkarni}, {Lambert}, {Le Guillou}, {Levine}, {Liang}, {Lim}, {Lintott},
  {Lupton}, {Mahabal}, {Marshall}, {Marshall}, {May}, {McKercher}, {Migliore},
  {Miller}, {Mills}, {Monet}, {Moniez}, {Neill}, {Nief}, {Nomerotski},
  {Nordby}, {O'Connor}, {Oliver}, {Olivier}, {Olsen}, {Ortiz}, {Owen}, {Pain},
  {Peterson}, {Petry}, {Pierfederici}, {Pietrowicz}, {Pike}, {Pinto}, {Plante},
  {Plate}, {Price}, {Prouza}, {Radeka}, {Rajagopal}, {Rasmussen}, {Regnault},
  {Ridgway}, {Ritz}, {Rosing}, {Roucelle}, {Rumore}, {Russo}, {Saha},
  {Sassolas}, {Schalk}, {Schindler}, {Schneider}, {Schumacher}, {Sebag},
  {Sembroski}, {Seppala}, {Shipsey}, {Silvestri}, {Smith}, {Smith}, {Strauss},
  {Stubbs}, {Sweeney}, {Szalay}, {Takacs}, {Thaler}, {Van Berg}, {Vanden Berk},
  {Vetter}, {Virieux}, {Xin}, {Walkowicz}, {Walter}, {Wang}, {Warner},
  {Willman}, {Wittman}, {Wolff}, {Wood-Vasey}, {Yoachim}, {Zhan}, \& {for the
  LSST Collaboration}}]{Ivezic-2008}
{Ivezic}, Z., {Tyson}, J.~A., {Abel}, B., {et~al.} 2008, \JournalTitle{ArXiv
  e-prints}, \href{http://arxiv.org/abs/0805.2366v4}{{\sffamily
  arXiv:0805.2366v4}}

\bibitem[{{Johnson} {et~al.}(2017){Johnson}, {Kochanek}, \&
  {Adams}}]{Johnson-2017}
{Johnson}, S.~A., {Kochanek}, C.~S., \& {Adams}, S.~M. 2017,
  \JournalTitle{ArXiv e-prints},
  \href{http://arxiv.org/abs/1712.03957}{{\sffamily arXiv:1712.03957
  [astro-ph.SR]}}

\bibitem[{Jones {et~al.}(2001--)Jones, Oliphant, Peterson, {et~al.}}]{scipy}
Jones, E., Oliphant, T., Peterson, P., {et~al.} 2001--, {SciPy}: Open source
  scientific tools for {Python}, http://www.scipy.org/

\bibitem[{{Jurcevic} {et~al.}(2000){Jurcevic}, {Pierce}, \&
  {Jacoby}}]{Jurcevic-2000}
{Jurcevic}, J.~S., {Pierce}, M.~J., \& {Jacoby}, G.~H. 2000,
  \href{http://dx.doi.org/10.1046/j.1365-8711.2000.03292.x}{\JournalTitle{\mnras},
  313, 868}

\bibitem[{{Kaiser} {et~al.}(2010){Kaiser}, {Burgett}, {Chambers}, {Denneau},
  {Heasley}, {Jedicke}, {Magnier}, {Morgan}, {Onaka}, \& {Tonry}}]{Kaiser-2010}
{Kaiser}, N., {Burgett}, W., {Chambers}, K., {et~al.} 2010,
  \href{http://dx.doi.org/10.1117/12.859188}{in \procspie, Vol. 7733,
  Ground-based and Airborne Telescopes III}, 77330E

\bibitem[{{Kiss} {et~al.}(2006){Kiss}, {Szab{\'o}}, \& {Bedding}}]{Kiss-2006}
{Kiss}, L.~L., {Szab{\'o}}, G.~M., \& {Bedding}, T.~R. 2006,
  \href{http://dx.doi.org/10.1111/j.1365-2966.2006.10973.x}{\JournalTitle{\mnras},
  372, 1721}

\bibitem[{{Kochanek} {et~al.}(2017){Kochanek}, {Fraser}, {Adams}, {Sukhbold},
  {Prieto}, {M{\"u}ller}, {Bock}, {Brown}, {Dong}, {Holoien}, {Khan},
  {Shappee}, \& {Stanek}}]{Kochanek-2017}
{Kochanek}, C.~S., {Fraser}, M., {Adams}, S.~M., {et~al.} 2017,
  \href{http://dx.doi.org/10.1093/mnras/stx291}{\JournalTitle{\mnras}, 467,
  3347}

\bibitem[{{Langer} \& {Maeder}(1995)}]{Langer-1995}
{Langer}, N., \& {Maeder}, A. 1995, \JournalTitle{\aap}, 295, 685

\bibitem[{{Law} {et~al.}(2009){Law}, {Kulkarni}, {Dekany}, {Ofek}, {Quimby},
  {Nugent}, {Surace}, {Grillmair}, {Bloom}, {Kasliwal}, {Bildsten}, {Brown},
  {Cenko}, {Ciardi}, {Croner}, {Djorgovski}, {van Eyken}, {Filippenko}, {Fox},
  {Gal-Yam}, {Hale}, {Hamam}, {Helou}, {Henning}, {Howell}, {Jacobsen},
  {Laher}, {Mattingly}, {McKenna}, {Pickles}, {Poznanski}, {Rahmer}, {Rau},
  {Rosing}, {Shara}, {Smith}, {Starr}, {Sullivan}, {Velur}, {Walters}, \&
  {Zolkower}}]{Law-2009}
{Law}, N.~M., {Kulkarni}, S.~R., {Dekany}, R.~G., {et~al.} 2009,
  \href{http://dx.doi.org/10.1086/648598}{\JournalTitle{\pasp}, 121, 1395}

\bibitem[{{Law} {et~al.}(2012){Law}, {Kraus}, {Street}, {Fulton},
  {Hillenbrand}, {Shporer}, {Lister}, {Baranec}, {Bloom}, {Bui}, {Burse},
  {Cenko}, {Das}, {Davis}, {Dekany}, {Filippenko}, {Kasliwal}, {Kulkarni},
  {Nugent}, {Ofek}, {Poznanski}, {Quimby}, {Ramaprakash}, {Riddle},
  {Silverman}, {Sivanandam}, \& {Tendulkar}}]{Law-2012}
{Law}, N.~M., {Kraus}, A.~L., {Street}, R., {et~al.} 2012,
  \href{http://dx.doi.org/10.1088/0004-637X/757/2/133}{\JournalTitle{\apj},
  757, 133}

\bibitem[{{Lee} {et~al.}(2012){Lee}, {Riffeser}, {Koppenhoefer}, {Seitz},
  {Bender}, {Hopp}, {G{\"o}ssl}, {Saglia}, {Snigula}, {Sweeney}, {Burgett},
  {Chambers}, {Grav}, {Heasley}, {Hodapp}, {Kaiser}, {Magnier}, {Morgan},
  {Price}, {Stubbs}, {Tonry}, \& {Wainscoat}}]{Lee-2012}
{Lee}, C.-H., {Riffeser}, A., {Koppenhoefer}, J., {et~al.} 2012,
  \href{http://dx.doi.org/10.1088/0004-6256/143/4/89}{\JournalTitle{\aj}, 143,
  89}

\bibitem[{{Levesque} {et~al.}(2005){Levesque}, {Massey}, {Olsen}, {Plez},
  {Josselin}, {Maeder}, \& {Meynet}}]{Levesque-2005}
{Levesque}, E.~M., {Massey}, P., {Olsen}, K.~A.~G., {et~al.} 2005,
  \href{http://dx.doi.org/10.1086/430901}{\JournalTitle{\apj}, 628, 973}

\bibitem[{{Levesque} {et~al.}(2006){Levesque}, {Massey}, {Olsen}, {Plez},
  {Meynet}, \& {Maeder}}]{Levesque-2006}
---. 2006, \href{http://dx.doi.org/10.1086/504417}{\JournalTitle{\apj}, 645,
  1102}

\bibitem[{{Lomb}(1976)}]{Lomb-1976}
{Lomb}, N.~R. 1976,
  \href{http://dx.doi.org/10.1007/BF00648343}{\JournalTitle{\apss}, 39, 447}

\bibitem[{{Maeder} \& {Meynet}(2000)}]{Maeder-2000}
{Maeder}, A., \& {Meynet}, G. 2000,
  \href{http://dx.doi.org/10.1146/annurev.astro.38.1.143}{\JournalTitle{\araa},
  38, 143}

\bibitem[{{Martins} \& {Palacios}(2013)}]{Martins-2013}
{Martins}, F., \& {Palacios}, A. 2013,
  \href{http://dx.doi.org/10.1051/0004-6361/201322480}{\JournalTitle{\aap},
  560, A16}

\bibitem[{{Masci} {et~al.}(2017){Masci}, {Laher}, {Rebbapragada}, {Doran},
  {Miller}, {Bellm}, {Kasliwal}, {Ofek}, {Surace}, {Shupe}, {Grillmair},
  {Jackson}, {Barlow}, {Yan}, {Cao}, {Cenko}, {Storrie-Lombardi}, {Helou},
  {Prince}, \& {Kulkarni}}]{Masci-2016}
{Masci}, F.~J., {Laher}, R.~R., {Rebbapragada}, U.~D., {et~al.} 2017,
  \href{http://dx.doi.org/10.1088/1538-3873/129/971/014002}{\JournalTitle{\pasp},
  129, 014002}

\bibitem[{{Massey}(1998)}]{Massey-1998}
{Massey}, P. 1998, \href{http://dx.doi.org/10.1086/305818}{\JournalTitle{\apj},
  501, 153}

\bibitem[{{Massey}(2003)}]{Massey-2003}
---. 2003,
  \href{http://dx.doi.org/10.1146/annurev.astro.41.071601.170033}{\JournalTitle{\araa},
  41, 15}

\bibitem[{{Massey}(2013)}]{Massey-2013}
---. 2013,
  \href{http://dx.doi.org/10.1016/j.newar.2013.05.002}{\JournalTitle{\nar}, 57,
  14}

\bibitem[{{Massey} \& {Evans}(2016)}]{Massey-2016}
{Massey}, P., \& {Evans}, K.~A. 2016,
  \href{http://dx.doi.org/10.3847/0004-637X/826/2/224}{\JournalTitle{\apj},
  826, 224}

\bibitem[{{Massey} {et~al.}(2017){Massey}, {Neugent}, \&
  {Levesque}}]{Massey-2017}
{Massey}, P., {Neugent}, K.~F., \& {Levesque}, E.~M. 2017,
  \href{http://dx.doi.org/10.1098/rsta.2016.0267}{\JournalTitle{Philosophical
  Transactions of the Royal Society of London Series A}, 375, 20160267}

\bibitem[{{Massey} {et~al.}(2007){Massey}, {Olsen}, {Hodge}, {Jacoby},
  {McNeill}, {Smith}, \& {Strong}}]{Massey-2007}
{Massey}, P., {Olsen}, K.~A.~G., {Hodge}, P.~W., {et~al.} 2007,
  \href{http://dx.doi.org/10.1086/513319}{\JournalTitle{\aj}, 133, 2393}

\bibitem[{{Massey} {et~al.}(2006){Massey}, {Olsen}, {Hodge}, {Strong},
  {Jacoby}, {Schlingman}, \& {Smith}}]{Massey-2006}
---. 2006, \href{http://dx.doi.org/10.1086/503256}{\JournalTitle{\aj}, 131,
  2478}

\bibitem[{{Massey} {et~al.}(2009){Massey}, {Silva}, {Levesque}, {Plez},
  {Olsen}, {Clayton}, {Meynet}, \& {Maeder}}]{Massey-2009}
{Massey}, P., {Silva}, D.~R., {Levesque}, E.~M., {et~al.} 2009,
  \href{http://dx.doi.org/10.1088/0004-637X/703/1/420}{\JournalTitle{\apj},
  703, 420}

\bibitem[{{Mauron} \& {Josselin}(2011)}]{Mauron-2011}
{Mauron}, N., \& {Josselin}, E. 2011,
  \href{http://dx.doi.org/10.1051/0004-6361/201013993}{\JournalTitle{\aap},
  526, A156}

\bibitem[{McKinney(2010)}]{pandas}
McKinney, W. 2010, in Proceedings of the 9th Python in Science Conference, ed.
  S.~van~der Walt \& J.~Millman, 51

\bibitem[{{Moriya} \& {Langer}(2015)}]{Moriya-2015}
{Moriya}, T.~J., \& {Langer}, N. 2015,
  \href{http://dx.doi.org/10.1051/0004-6361/201424957}{\JournalTitle{\aap},
  573, A18}

\bibitem[{{Morozova} {et~al.}(2017){Morozova}, {Piro}, \&
  {Valenti}}]{Morozova-2017}
{Morozova}, V., {Piro}, A.~L., \& {Valenti}, S. 2017, \JournalTitle{ArXiv
  e-prints}, \href{http://arxiv.org/abs/1709.04928}{{\sffamily arXiv:1709.04928
  [astro-ph.HE]}}

\bibitem[{{Mould}(1987)}]{Mould-1987}
{Mould}, J. 1987, \href{http://dx.doi.org/10.1086/132091}{\JournalTitle{\pasp},
  99, 1127}

\bibitem[{{Neilson} \& {Lester}(2008)}]{Neilson-2008}
{Neilson}, H.~R., \& {Lester}, J.~B. 2008,
  \href{http://dx.doi.org/10.1086/588650}{\JournalTitle{\apj}, 684, 569}

\bibitem[{{Ofek} {et~al.}(2012){Ofek}, {Laher}, {Law}, {Surace}, {Levitan},
  {Sesar}, {Horesh}, {Poznanski}, {van Eyken}, {Kulkarni}, {Nugent},
  {Zolkower}, {Walters}, {Sullivan}, {Ag{\"u}eros}, {Bildsten}, {Bloom},
  {Cenko}, {Gal-Yam}, {Grillmair}, {Helou}, {Kasliwal}, \&
  {Quimby}}]{Ofek-2012}
{Ofek}, E.~O., {Laher}, R., {Law}, N., {et~al.} 2012,
  \href{http://dx.doi.org/10.1086/664065}{\JournalTitle{\pasp}, 124, 62}

\bibitem[{{Oppermann} {et~al.}(2013){Oppermann}, {Selig}, {Bell}, \&
  {En{\ss}lin}}]{Oppermann-2013}
{Oppermann}, N., {Selig}, M., {Bell}, M.~R., \& {En{\ss}lin}, T.~A. 2013,
  \href{http://dx.doi.org/10.1103/PhysRevE.87.032136}{\JournalTitle{\pre}, 87,
  032136}

\bibitem[{{Paxton} {et~al.}(2011){Paxton}, {Bildsten}, {Dotter}, {Herwig},
  {Lesaffre}, \& {Timmes}}]{Mesa-1}
{Paxton}, B., {Bildsten}, L., {Dotter}, A., {et~al.} 2011,
  \href{http://dx.doi.org/10.1088/0067-0049/192/1/3}{\JournalTitle{\apjs}, 192,
  3}

\bibitem[{{Paxton} {et~al.}(2013){Paxton}, {Cantiello}, {Arras}, {Bildsten},
  {Brown}, {Dotter}, {Mankovich}, {Montgomery}, {Stello}, {Timmes}, \&
  {Townsend}}]{Mesa-2}
{Paxton}, B., {Cantiello}, M., {Arras}, P., {et~al.} 2013,
  \href{http://dx.doi.org/10.1088/0067-0049/208/1/4}{\JournalTitle{\apjs}, 208,
  4}

\bibitem[{{Paxton} {et~al.}(2015){Paxton}, {Marchant}, {Schwab}, {Bauer},
  {Bildsten}, {Cantiello}, {Dessart}, {Farmer}, {Hu}, {Langer}, {Townsend},
  {Townsley}, \& {Timmes}}]{Mesa-3}
{Paxton}, B., {Marchant}, P., {Schwab}, J., {et~al.} 2015,
  \href{http://dx.doi.org/10.1088/0067-0049/220/1/15}{\JournalTitle{\apjs},
  220, 15}

\bibitem[{{Paxton} {et~al.}(2018){Paxton}, {Schwab}, {Bauer}, {Bildsten},
  {Blinnikov}, {Duffell}, {Farmer}, {Goldberg}, {Marchant}, {Sorokina},
  {Thoul}, {Townsend}, \& {Timmes}}]{Mesa-4}
{Paxton}, B., {Schwab}, J., {Bauer}, E.~B., {et~al.} 2018,
  \href{http://dx.doi.org/10.3847/1538-4365/aaa5a8}{\JournalTitle{\apjs}, 234,
  34}

\bibitem[{{Percy} \& {Abachi}(2013)}]{Percy-2013}
{Percy}, J.~R., \& {Abachi}, R. 2013, \JournalTitle{Journal of the American
  Association of Variable Star Observers (JAAVSO)}, 41, 193

\bibitem[{{Pierce} {et~al.}(2000){Pierce}, {Jurcevic}, \&
  {Crabtree}}]{Pierce-2000}
{Pierce}, M.~J., {Jurcevic}, J.~S., \& {Crabtree}, D. 2000,
  \href{http://dx.doi.org/10.1046/j.1365-8711.2000.03196.x}{\JournalTitle{\mnras},
  313, 271}

\bibitem[{{Pojmanski}(2002)}]{Pojmanski-2002}
{Pojmanski}, G. 2002, \JournalTitle{\actaa}, 52, 397

\bibitem[{{Rasmussen} \& {Williams}(2005)}]{Rasmussen-2005}
{Rasmussen}, C., \& {Williams}, C. 2005, Gaussian Processes for Machine
  Learning (Adaptive Computation and Machine Learning) (Cambridge, MA: MIT
  Press)

\bibitem[{{Rau} {et~al.}(2009){Rau}, {Kulkarni}, {Law}, {Bloom}, {Ciardi},
  {Djorgovski}, {Fox}, {Gal-Yam}, {Grillmair}, {Kasliwal}, {Nugent}, {Ofek},
  {Quimby}, {Reach}, {Shara}, {Bildsten}, {Cenko}, {Drake}, {Filippenko},
  {Helfand}, {Helou}, {Howell}, {Poznanski}, \& {Sullivan}}]{Rau-2009}
{Rau}, A., {Kulkarni}, S.~R., {Law}, N.~M., {et~al.} 2009,
  \href{http://dx.doi.org/10.1086/605911}{\JournalTitle{\pasp}, 121, 1334}

\bibitem[{{Riess} {et~al.}(2012){Riess}, {Fliri}, \&
  {Valls-Gabaud}}]{Riess-2012}
{Riess}, A.~G., {Fliri}, J., \& {Valls-Gabaud}, D. 2012,
  \href{http://dx.doi.org/10.1088/0004-637X/745/2/156}{\JournalTitle{\apj},
  745, 156}

\bibitem[{{Riffeser} {et~al.}(2001){Riffeser}, {Fliri}, {G{\"o}ssl}, {Bender},
  {Hopp}, {B{\"a}rnbantner}, {Ries}, {Barwig}, {Seitz}, \&
  {Mitsch}}]{Riffeser-2001}
{Riffeser}, A., {Fliri}, J., {G{\"o}ssl}, C.~A., {et~al.} 2001,
  \href{http://dx.doi.org/10.1051/0004-6361:20011288}{\JournalTitle{\aap}, 379,
  362}

\bibitem[{{Saha} \& {Vivas}(2017)}]{Saha-2017}
{Saha}, A., \& {Vivas}, A.~K. 2017, \JournalTitle{ArXiv e-prints},
  \href{http://arxiv.org/abs/1709.10156}{{\sffamily arXiv:1709.10156
  [astro-ph.IM]}}

\bibitem[{{Scargle}(1982)}]{Scargle-1982}
{Scargle}, J.~D. 1982,
  \href{http://dx.doi.org/10.1086/160554}{\JournalTitle{\apj}, 263, 835}

\bibitem[{{Schwarzenberg-Czerny}(1989)}]{Schwarzenberg-1989}
{Schwarzenberg-Czerny}, A. 1989,
  \href{http://dx.doi.org/10.1093/mnras/241.2.153}{\JournalTitle{\mnras}, 241,
  153}

\bibitem[{{Selig} {et~al.}(2013){Selig}, {Bell}, {Junklewitz}, {Oppermann},
  {Reinecke}, {Greiner}, {Pachajoa}, \& {En{\ss}lin}}]{Selig-2013}
{Selig}, M., {Bell}, M.~R., {Junklewitz}, H., {et~al.} 2013,
  \href{http://dx.doi.org/10.1051/0004-6361/201321236}{\JournalTitle{\aap},
  554, A26}

\bibitem[{{Shiode} \& {Quataert}(2014)}]{Shiode-2014}
{Shiode}, J.~H., \& {Quataert}, E. 2014,
  \href{http://dx.doi.org/10.1088/0004-637X/780/1/96}{\JournalTitle{\apj}, 780,
  96}

\bibitem[{{Smartt}(2009)}]{Smartt-1}
{Smartt}, S.~J. 2009,
  \href{http://dx.doi.org/10.1146/annurev-astro-082708-101737}{\JournalTitle{\araa},
  47, 63}

\bibitem[{{Smartt}(2015)}]{Smartt-2}
---. 2015, \href{http://dx.doi.org/10.1017/pasa.2015.17}{\JournalTitle{\pasa},
  32, e016}

\bibitem[{{Smartt} {et~al.}(2009){Smartt}, {Eldridge}, {Crockett}, \&
  {Maund}}]{Smartt-2009}
{Smartt}, S.~J., {Eldridge}, J.~J., {Crockett}, R.~M., \& {Maund}, J.~R. 2009,
  \href{http://dx.doi.org/10.1111/j.1365-2966.2009.14506.x}{\JournalTitle{\mnras},
  395, 1409}

\bibitem[{{Smith} {et~al.}(2011){Smith}, {Li}, {Filippenko}, \&
  {Chornock}}]{Nathan-2011}
{Smith}, N., {Li}, W., {Filippenko}, A.~V., \& {Chornock}, R. 2011,
  \href{http://dx.doi.org/10.1111/j.1365-2966.2011.17229.x}{\JournalTitle{\mnras},
  412, 1522}

\bibitem[{{Soraisam} {et~al.}(2017){Soraisam}, {Gilfanov}, {Kupfer}, {Masci},
  {Shafter}, {Prince}, {Kulkarni}, {Ofek}, \& {Bellm}}]{Soraisam-2017}
{Soraisam}, M.~D., {Gilfanov}, M., {Kupfer}, T., {et~al.} 2017,
  \href{http://dx.doi.org/10.1051/0004-6361/201629368}{\JournalTitle{\aap},
  599, A48}

\bibitem[{{Stellingwerf}(1978)}]{Stellingwerf-1978}
{Stellingwerf}, R.~F. 1978,
  \href{http://dx.doi.org/10.1086/156444}{\JournalTitle{\apj}, 224, 953}

\bibitem[{{Stetson}(1987)}]{Stetson}
{Stetson}, P.~B. 1987,
  \href{http://dx.doi.org/10.1086/131977}{\JournalTitle{\pasp}, 99, 191}

\bibitem[{{Stetson}(1990)}]{Stetson-1990}
---. 1990, \href{http://dx.doi.org/10.1086/132719}{\JournalTitle{\pasp}, 102,
  932}

\bibitem[{{Stothers}(1969)}]{Stothers-1969}
{Stothers}, R. 1969,
  \href{http://dx.doi.org/10.1086/149987}{\JournalTitle{\apj}, 156, 541}

\bibitem[{{Sugimoto} \& {Nomoto}(1974)}]{Sugimoto}
{Sugimoto}, D., \& {Nomoto}, K.-I. 1974, in IAU Symposium, Vol.~66, Late Stages
  of Stellar Evolution, ed. R.~J. {Tayler} \& J.~E. {Hesser}, 105

\bibitem[{{Sukhbold} {et~al.}(2016){Sukhbold}, {Ertl}, {Woosley}, {Brown}, \&
  {Janka}}]{Sukhbold-2016}
{Sukhbold}, T., {Ertl}, T., {Woosley}, S.~E., {Brown}, J.~M., \& {Janka}, H.-T.
  2016,
  \href{http://dx.doi.org/10.3847/0004-637X/821/1/38}{\JournalTitle{\apj}, 821,
  38}

\bibitem[{{Townsend} \& {Teitler}(2013)}]{Townsend-2013}
{Townsend}, R.~H.~D., \& {Teitler}, S.~A. 2013,
  \href{http://dx.doi.org/10.1093/mnras/stt1533}{\JournalTitle{\mnras}, 435,
  3406}

\bibitem[{van~der Walt {et~al.}(2011)van~der Walt, Colbert, \&
  Varoquaux}]{numpy}
van~der Walt, S., Colbert, S.~C., \& Varoquaux, G. 2011,
  \href{http://dx.doi.org/10.1109/MCSE.2011.37}{\JournalTitle{Computing In
  Science \& Engineering}, 13, 22}

\bibitem[{{van Dyk} {et~al.}(2017){van Dyk}, {Filippenko}, {Fox}, {Kelly},
  {Milisavljevic}, \& {Smith}}]{van-Dyk-2017}
{van Dyk}, S.~D., {Filippenko}, A.~V., {Fox}, O.~D., {et~al.} 2017,
  \JournalTitle{The Astronomer's Telegram}, 10378

\bibitem[{{Van Dyk} {et~al.}(2003){Van Dyk}, {Li}, \&
  {Filippenko}}]{Van-Dyk-2003}
{Van Dyk}, S.~D., {Li}, W., \& {Filippenko}, A.~V. 2003,
  \href{http://dx.doi.org/10.1086/345748}{\JournalTitle{\pasp}, 115, 1}

\bibitem[{{Vink} {et~al.}(2001){Vink}, {de Koter}, \& {Lamers}}]{Vink01}
{Vink}, J.~S., {de Koter}, A., \& {Lamers}, H.~J.~G.~L.~M. 2001,
  \href{http://dx.doi.org/10.1051/0004-6361:20010127}{\JournalTitle{\aap}, 369,
  574}

\bibitem[{{Wang} {et~al.}(2012){Wang}, {Khardon}, \& {Protopapas}}]{Wang-2012}
{Wang}, Y., {Khardon}, R., \& {Protopapas}, P. 2012,
  \href{http://dx.doi.org/10.1088/0004-637X/756/1/67}{\JournalTitle{\apj}, 756,
  67}

\bibitem[{{Wood} {et~al.}(1983){Wood}, {Bessell}, \& {Fox}}]{Wood-1983}
{Wood}, P.~R., {Bessell}, M.~S., \& {Fox}, M.~W. 1983,
  \href{http://dx.doi.org/10.1086/161265}{\JournalTitle{\apj}, 272, 99}

\bibitem[{{Yang} \& {Jiang}(2011)}]{Yang-2011}
{Yang}, M., \& {Jiang}, B.~W. 2011,
  \href{http://dx.doi.org/10.1088/0004-637X/727/1/53}{\JournalTitle{\apj}, 727,
  53}

\bibitem[{{Yang} \& {Jiang}(2012)}]{Yang-2012}
---. 2012,
  \href{http://dx.doi.org/10.1088/0004-637X/754/1/35}{\JournalTitle{\apj}, 754,
  35}

\bibitem[{{Yoon} \& {Cantiello}(2010)}]{Yoon-2010}
{Yoon}, S.-C., \& {Cantiello}, M. 2010,
  \href{http://dx.doi.org/10.1088/2041-8205/717/1/L62}{\JournalTitle{\apjl},
  717, L62}

\end{thebibliography}

\begin{table*}[ht]
\caption{Properties of the RSGs in M31\tablenotemark{a}}\label{table_per}
\renewcommand\arraystretch{1.0}
\renewcommand{\tabcolsep}{15pt}
\centering
\begin{tabularx}{1.0\textwidth}{*{7}l*{1}{X}}
\hline
ID\tablenotemark{b}	
&Star\tablenotemark{c}		
&$M_{K}$\tablenotemark{c} 
&$T_{eff}$ (K)\tablenotemark{c}		&$\log(L/L_{\odot})$\tablenotemark{c}
&$\left<m_{R}\right>$
&$\Delta m_{R}$\tablenotemark{d}
&Period (days)\\
\hline
1    &J003950.86+405332.0    &-10.03    &3850.0    &4.86    &18.63    &0.27    &391\\
2    &J003950.98+405422.5    &-10.28    &3650.0    &4.89    &18.46    &0.11    & --\\
3    &J003957.00+410114.6    &-9.4    &3650.0    &4.54    &18.80    &0.12    & --\\
4    &J004015.18+405947.7    &-10.22    &3700.0    &4.89    &18.62    &0.23    &391\\
5    &J004015.86+405514.1    &-8.86    &3950.0    &4.42    &18.74    &(0.08)    & --\\
6    &J004019.15+404150.8    &-9.87    &3750.0    &4.76    &18.72    &0.10    & --\\
\hline
\end{tabularx}
\begin{flushleft}
\tablenotetext{a}{Measurements not available are indicated by a dash.}
\tablenotetext{b}{Our ID follows the order in which the RSGs appear in the ME16 catalog.}
\tablenotetext{c}{These values are obtained from ME16.}
\tablenotetext{d}{Bracketed value is for RSG whose variability is below our threshold (see Fig.~\ref{fig:static}).}
\tablenotetext{}{(This is only a part of the table; the full version is available in machine-readable form online.)}
\end{flushleft}
\end{table*}

\end{document}